\documentclass[twocolumn,aps,prl]{revtex4-1}

\usepackage[T1]{fontenc}
\usepackage[utf8]{inputenc}
\usepackage{color}
\usepackage{graphicx}
\usepackage{amsmath}
\usepackage[english]{babel}
\usepackage{url}
\usepackage{braket}
\usepackage{subfig}
\usepackage{float}
\usepackage{verbatim}

\graphicspath{{files/}}

\makeatletter

\@ifundefined{textcolor}{}
{%
 \definecolor{BLACK}{gray}{0}
 \definecolor{GRAY}{gray}{0.50}
 \definecolor{WHITE}{gray}{1}
 \definecolor{RED}{rgb}{1,0,0}
 \definecolor{GREEN}{rgb}{0,0.50,0}
 \definecolor{BLUE}{rgb}{0,0,1}
 \definecolor{CYAN}{cmyk}{1,0,0,0}
 \definecolor{MAGENTA}{cmyk}{0,1,0,0}
 \definecolor{YELLOW}{cmyk}{0,0,1,0}
}

\makeatother

\begin{document}

\title{Variational study of the flux tube recombination in the two quarks and two quarks system
i Lattice QCD}

\author{M. Cardoso} \email{mjdcc@cftp.ist.utl.pt}

\author{N. Cardoso} \email{nunocardoso@cftp.ist.utl.pt}

\author{P. Bicudo} \email{bicudo@ist.utl.pt}

\affiliation{CFTP, Departamento de Física, Instituto Superior Técnico (Universidade
Técnica de Lisboa), Av. Rovisco Pais, 1049-001 Lisboa, Portugal}

\begin{abstract}
The color fields in a system composed by two static quarks and two
static antiquarks are studied.
In particular, we consider the four particles in the corners of a rectangle,
and two possible alignment of the particles, one in which the quarks
are at the same side of the rectangle, and the other where they are at opposite sides.
We use a variational method, to probe not only the ground state but also the first
excited state. This results permit us to observe and interpret the flux-tube recombination in the
mesons to mesons and the tetraquark to mesons transitions, for both states.
The results are compared with previous results for the static potential and the Casimir scaling predictions.
\end{abstract}
\maketitle

\section{Introduction}

Systems constituted by two quarks and two antiquarks are of extreme importance for strong interaction physics.
Not only because they are a starting point for meson-meson scattering,
but also because of the possible existence of bound-states --- tetraquarks, initially predicted by Jaffe \cite{Jaffe:1976ig}.
There are several observed resonances which are candidates to tetraquarks \cite{Choi:2003ue, Acosta:2003zx, :2009xt}.
The most recent are the $Z_b^+$ particles reported by the Belle collaboration
\cite{Collaboration:2011gja,Belle:2011aa}.
While it remains difficult to study tetraquarks in Lattice QCD with dynamical quarks
\cite{Wagner:2011ev,Prelovsek:2010zza}
the static tetraquark potentials and flux tubes have been studied in precise
lattice QCD simmulations \cite{Okiharu:2004ve,Alexandrou:2004ak,Cardoso:2011fq}.
The system of two quarks and two antiquarks in $SU(2)$ was also studied using a variational method \cite{Pennanen:1998nu}.

The Lattice QCD results for the static potential seem to confirm that the ground
state potential is well described by a generalized flip-flop potential.
This piecewise potential was already proposed and used by different authors
\cite{Lenz:1985jk,Oka:1985vg,Oka:1984yx,Miyazawa:1980ft,Miyazawa:1979vx,Karliner:2003dt,
Zouzou:1986qh,Deng:2012wi,Vijande:2007ix}.

as a device to eliminate the physically non existent long
distance Van der Waals interactions arising from the naive potential
models based on sum of two-body Casimir scaled potentials
\cite{Fishbane:1977ay,Appelquist:1978rt,Willey:1978fm,Matsuyama:1978hf,Gavela:1979zu,Feinberg:1983zz}.

\begin{figure}
	\centering
	\includegraphics[width=0.2\textwidth]{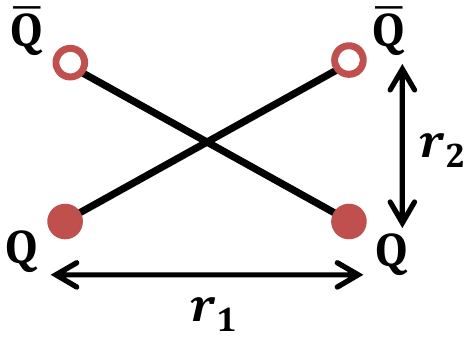}
	\includegraphics[width=0.2\textwidth]{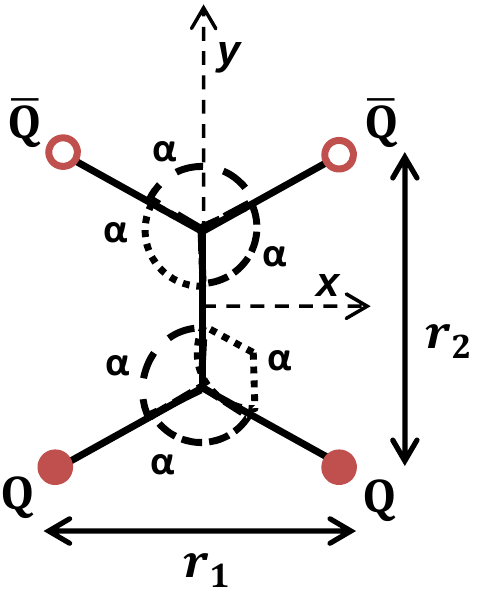}
	\caption{The shape of the minimal string which links the four particles in the tetraquark. Note that when the diquark and the diantiquark are close the five segment structure collapses into a four segment one.}
	\label{tetraq_string}
\end{figure}

The lattice studies indicate that the ground state is well described
by the generalized flip-flop potential
\begin{equation}
V_{FF}=\mbox{min}(V_{I},V_{II},V_{T}) \, .
\label{V_FF}
\end{equation}
This generalized potential differs from the firstly proposed flip-flop models because, besides the meson-meson domains already present in earlier models, it also includes a tetraquark domain.
$V_I$ and $V_{II}$, in Eq. (\ref{V_FF}), are the two possible two-meson potentials,
given as sum of the two intrameson potentials.
 Explicitly,
\begin{equation}
V_{I}=V_{M}(|\mathbf{r}_{1}-\mathbf{r}_{3}|)+V_{M}(|\mathbf{r}_{2}-\mathbf{r}_{4}|)\, ,
\end{equation}
and
\begin{equation}
V_{II}=V_{M}(|\mathbf{r}_{1}-\mathbf{r}_{4}|)+V_{M}(|\mathbf{r}_{2}-\mathbf{r}_{3}|)\, .
\end{equation}
where the intrameson potential $V_{M}$ is well fitted in the static
limit as a Cornell potential: $V_{M}(r)=C-\frac{\gamma}{r}+\sigma r$.
The potential $V_{T}$, is the tetraquark potential and is given according to \cite{Alexandrou:2004ak} and \cite{Okiharu:2004ve} by
\begin{equation}
V_{T}=C+\alpha\sum\frac{\lambda_{i}}{2}\cdot\frac{\lambda_{j}}{2}\frac{1}{r_{ij}}+\sigma L_{min}
\end{equation}
where $L_{min}$ is the minimal distance linking the four particles, as depicted 
in Fig. \ref{tetraq_string}.

 However, the value of the static potential alone, is not sufficient to understand the confinement in tetraquarks. Confinement can be researched both with the measurement of the color fields and with the study of the colour wave functions of the tetraquark. 
Recently we computed the color-electric and color-magnetic fields generated by a static tetraquark \cite{Cardoso:2011fq}.
We confirmed that for quark and antiquark distances compatible with a tetraquark potential, tetraquark color flux tubes actually are created. Thus the mechanism of confinement in static tetraquarks is the localization of the color fields in fundamental flux tubes.

But it remains important to clarify what are the color wave functions of static tetraquarks, and to observe the flip-flop of both the flux tubes and the wave functions of static teraquarks. Here we apply the variational principle to observe whether the flip-flop transition occurs for the color flux tubes in a quark-quark-antiquark-antiquark($QQ\bar{Q}\bar{Q}$) system.

In Section II, we discuss the two possible color singlets of the four-particle system, while 
in section III, we review the Wilson Loop of the tetraquark system.
Then, in section IV, we make the synthesis of the two previous sections and develop a variational method
to describe the $QQ\bar{Q}\bar{Q}$ system, either in a tetraquark color state or in a two meson state.
In section V, the method we use to compute the color fields in the lattice is described.
In section VI, this method is used to compute the color fields in the system, for different arrangements
of the four particles. In section VII our results are discussed and in section VIII the conclusions are
presented.

\section{A two component basis for the tetraquark color wavefunction}

\begin{figure}
	\centering
	\includegraphics[width=8cm]{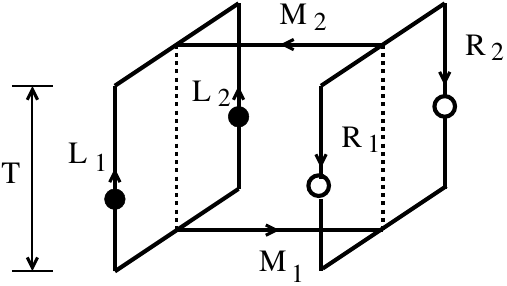}
	\caption{Wilson loop operator used by \cite{Okiharu:2004ve} and \cite{Alexandrou:2004ak}
	to calculate the static potential of the tetraquark system.}
	\label{tetraq_wloop}
\end{figure}

Notice a tetraquark must be a color singlet due to confinement and to gauge invariance, but there are two possible color wave functions for a color singlet tetraquark.
For instance, if the system is on the domain where $V_{FF}=V_{I}$ (domain I), we expect the color wavefunction of the ground state to be given by
\begin{equation}
|I\rangle=\frac{1}{3}\sum_{ij}|Q_{i}Q_{j}\bar{Q}_{i}\bar{Q}_{j}\rangle\,.
\end{equation}
If the system is on domain $II$, we expect the ground state to be given by
\begin{equation}
|II\rangle=\frac{1}{3}\sum_{ij}|Q_{i}Q_{j}\bar{Q}_{j}\bar{Q}_{i}\rangle\,.
\end{equation}
Note that these two systems are not orthogonal as $\langle I|II\rangle=\frac{1}{3}$.

When $V_{FF}=V_{T}$, we expect that the two quarks form an antitriplet and the two
antiquarks a triplet. So, the wavefunction is given by,
\begin{equation}
	|A\rangle = \mathcal{N} \epsilon_{ijk} | Q_i Q_j \rangle \epsilon_{klm} |\bar{Q}_l \bar{Q}_m \rangle\,,
\label{eq:Twf}
\end{equation}
where the $A$ stand for antisymmetric since this wavefunction is antisymmetric for the exchange of two quarks or two antiquarks. 
Eq. (\ref{eq:Twf}) can be simplified by contracting the tensors, and by imposing the normalization $\langle A | A \rangle = 1$,
\begin{equation}
|A\rangle=\frac{1}{2\sqrt{3}}(|Q_{i}Q_{j}\bar{Q}_{i}\bar{Q}_{j}\rangle-|Q_{i}Q_{j}\bar{Q}_{j}\bar{Q}_{i}\rangle)=\frac{\sqrt{3}}{2}(|I\rangle-|II\rangle)\,.
\end{equation}
We can also construct a color symmetric state as
\begin{equation}
|S\rangle=\frac{1}{2\sqrt{6}}(|Q_{i}Q_{j}\bar{Q}_{i}\bar{Q}_{j}\rangle+|Q_{i}Q_{j}\bar{Q}_{j}\bar{Q}_{i}\rangle)=\sqrt{\frac{3}{8}}(|I\rangle+|II\rangle)\,.
\label{eq_S}
\end{equation}
Any color single state of two quarks and two antiquarks can be decomposed in two different sets of basis vectors, either $|I\rangle$ and
$|II\rangle$ or $|A\rangle$ and $|S\rangle$. The second set
of kets has the advantage of forming an orthonormal basis. $|I\rangle$
and $|II\rangle$ can be written as 
\begin{equation}
|I\rangle=\sqrt{\frac{2}{3}}|S\rangle+\frac{1}{\sqrt{3}}|A\rangle\,,
\end{equation}
\begin{equation}
|II\rangle=\sqrt{\frac{2}{3}}|S\rangle-\frac{1}{\sqrt{3}}|A\rangle\,.
\end{equation}

\begin{figure}[!t]
\begin{centering}
    \subfloat[\label{fig:para}]{
\begin{centering}
    \includegraphics[width=4cm]{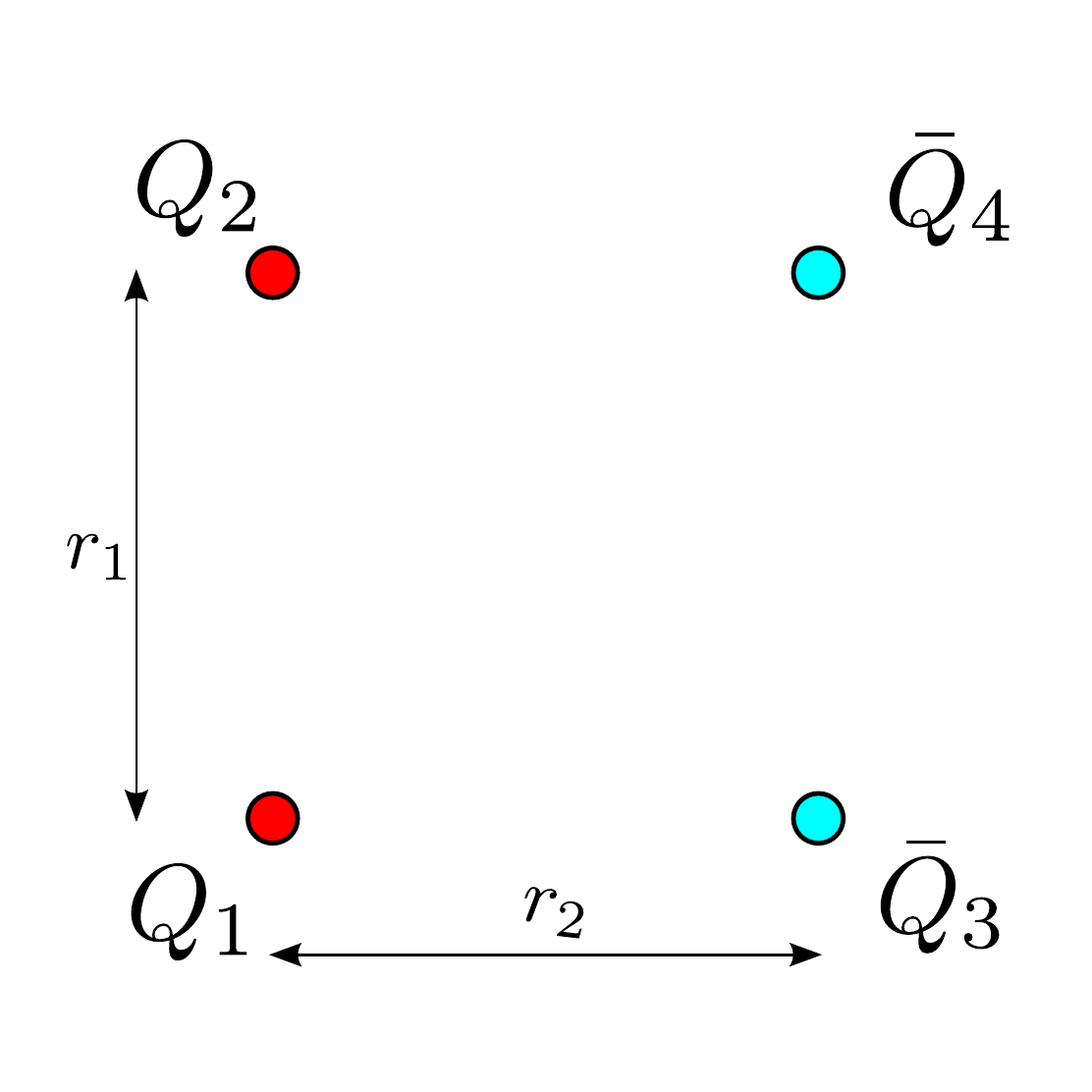}
\par\end{centering}}
    \subfloat[\label{fig:antip}]{
\begin{centering}
    \includegraphics[width=4cm]{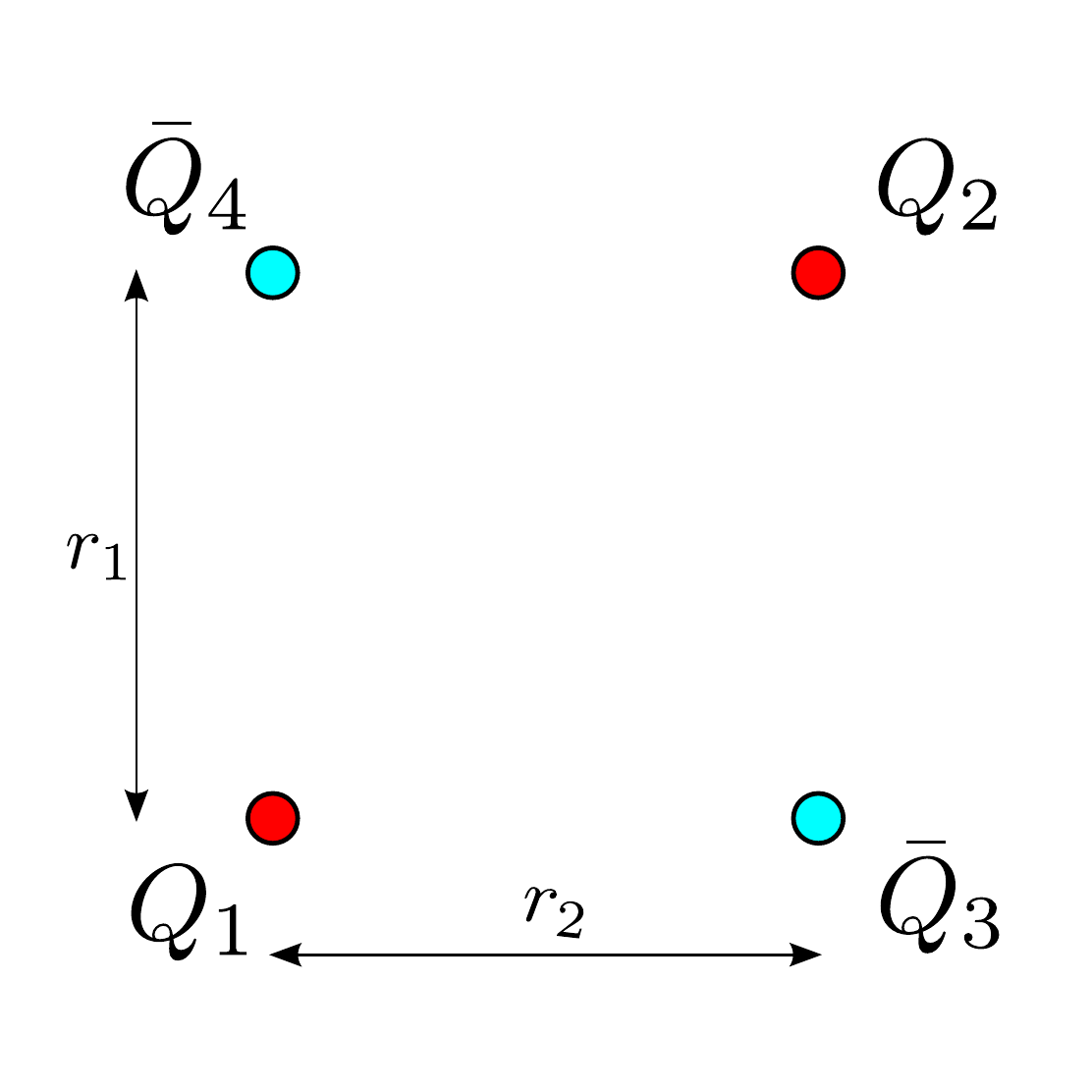}
\par\end{centering}}
\par\end{centering}
	\caption{\protect\subref{fig:para} Parallel alignment. \protect\subref{fig:antip} Antiparallel alignment.}
	\label{geom}
\end{figure}

Therefore, we can describe our state with a two component vector and the potential
itself as a two by two matrix. The lowest eigenvalue of the matrix
is corresponds to the static potential measured on the lattice, with
the corresponding eigenvector being $|I\rangle$, $|II\rangle$ or
$|A\rangle$ depending on the domain considered. Since the potential is
hermitian, the eigenvector of the excited state of the potential is
orthogonal to the one of the ground state. Therefore, these eigenvectors
must be 
\begin{equation}
|\bar{I}\rangle=-\frac{1}{\sqrt{3}}|S\rangle+\sqrt{\frac{2}{3}}|A\rangle\,,
\end{equation}
for the domain I, and
\begin{equation}
|\bar{II}\rangle=\frac{1}{\sqrt{3}}|S\rangle+\sqrt{\frac{2}{3}}|A\rangle\,,
\end{equation}
for the domain II.  In the tetraquark domain the excited state eigenvector must be $|S\rangle$, defined in  Eq. (\ref{eq_S}).

\section{The Tetraquark Wilson Loop}

\begin{figure}
	\centering
	\includegraphics[width=0.50\textwidth]{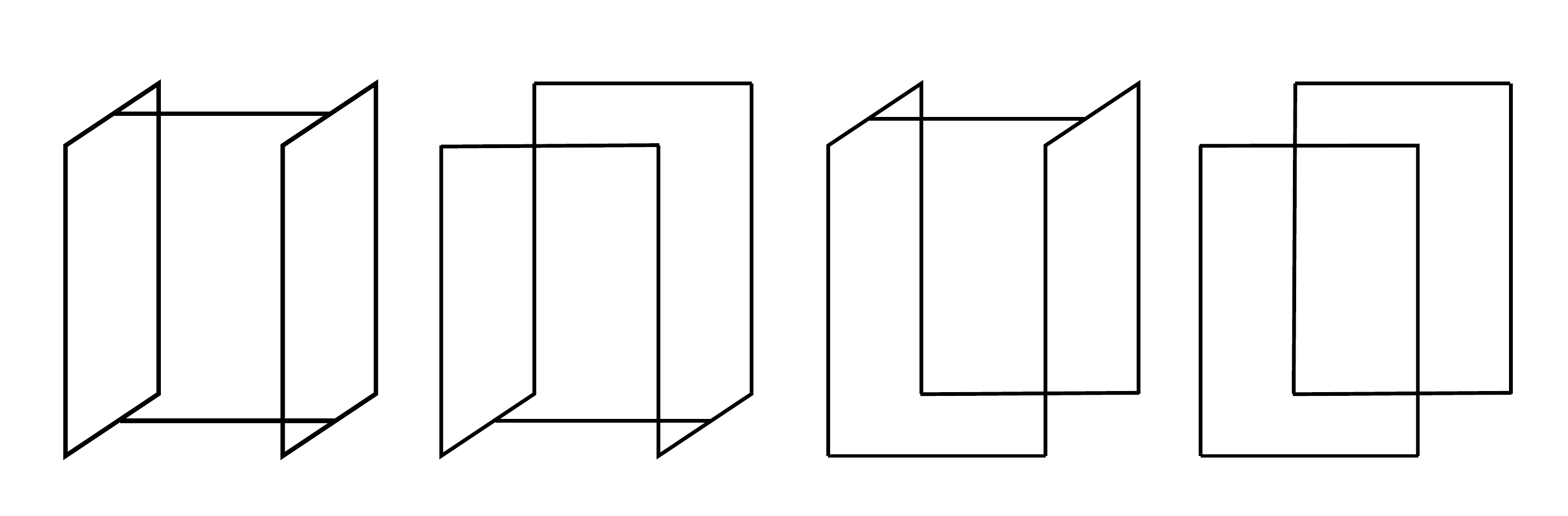}
	\caption{Elements of the Wilson loop matrix in the parallel alignment case.}
	\label{para_loop}
\end{figure}

The static potential for the tetraquark has been studied in the lattice by \cite{Alexandrou:2004ak} and \cite{Okiharu:2004ve}.
The Wilson loop operator for the tetraquark system, illustrated in Fig. \ref{tetraq_wloop}, is given by
\begin{equation}
	W_{4Q} = \frac{1}{3} \mbox{Tr}[ M_1 R_{12} M_2 L_{12} ] \,,
\end{equation}
with
\begin{eqnarray}
	R_{12}^{ii'} &=& \frac{1}{2} \epsilon_{ijk} \epsilon_{i'j'k'} R_1^{jj'} R_2^{kk'}\,, \\ \nonumber
	L_{12}^{ii'} &=& \frac{1}{2} \epsilon_{ijk} \epsilon_{i'j'k'} L_1^{jj'} L_2^{kk'} \, .
\end{eqnarray}

This Wilson Loop has the quantum numbers of color singlet system where
the two quarks form an antitriplet and the two antiquarks form a triplet.

This lattice studies, indicate that the static potential of the two-quark
and two-antiquark system is a generalized flip-flop potential,
\begin{equation}
	V_{FF} = \min( V_T , V_{M_1 M_2}, V_{M_3 M_4} )\,,
\end{equation}
where $V_{M_1 M_2}$ and $V_{M_3 M_4}$ are the two possible two-meson
potentials, given by the sum of two independent intra-meson potentials
$V_{M_1 M_2} = V_{M_1} + V_{M_2}$, and $V_T$ is the tetraquark potential
which corresponds to the sector where the four particles are confined,
linked by a single fundamental string.

For the special case where the four particles form as rectangle as in
Fig. \ref{tetraq_string}, $L_{min}$ is given by
$L_{min} = \sqrt{3} r_1 + r_2$ for $r_2 > \frac{r_1}{\sqrt{3}}$ 
\cite{Bicudo:2010mv,Bicudo:2008yr}. If we neglect the Coulomb part of $V_{FF}$, the flip-flop potential with the linear potentials only produces a tetraquark domain for distances
$r_2 \gtrsim \sqrt{3} r_1$.

\section{Variational Method for the $QQ\bar{Q}\bar{Q}$ system}

\begin{figure}
	\centering
	\includegraphics[width=0.50\textwidth]{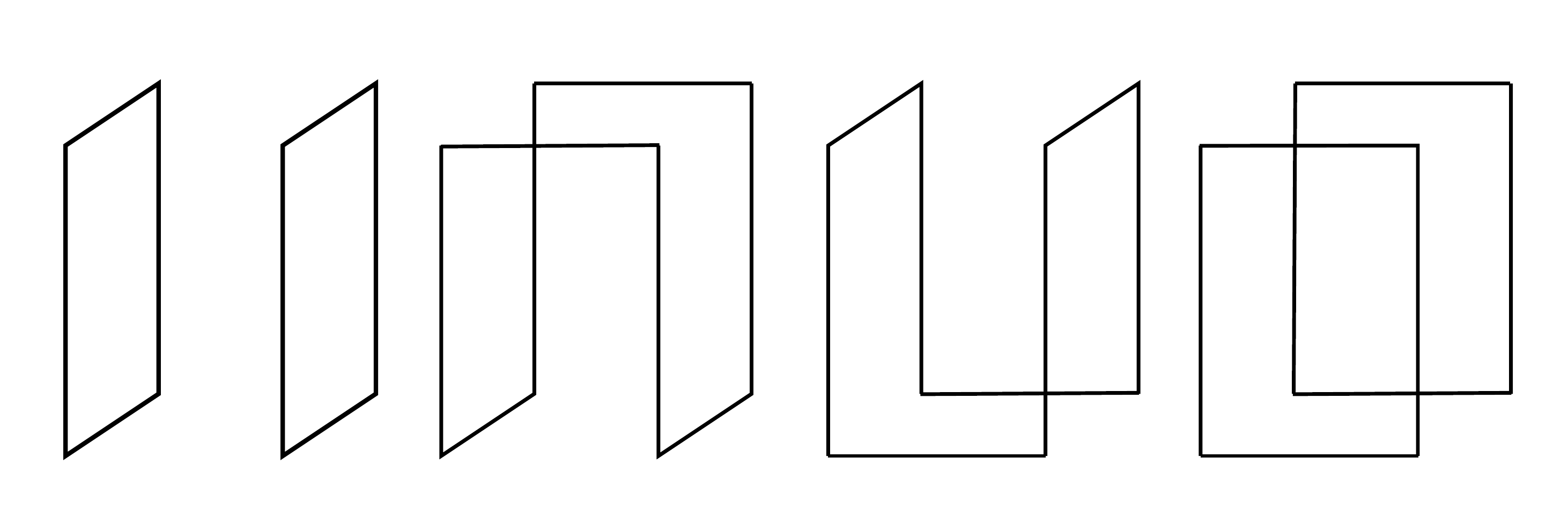}
	\caption{Elements of the Wilson loop matrix in the antiparallel alignment case.}
	\label{antip_loop}
\end{figure}

Now we extend the Wilson loop in order to obtain the true ground state of the tetraquark system and also to obtain the first excited state. 
To achieve this, we note that the Wilson Loop operator can be written as a correlation of a certain operator at different times $W(t)=\langle \hat{\mathcal{O}}(t) \hat{\mathcal{O}}^\dagger(0) \rangle$. This can be
generalized by considering instead a base of operators $\hat{\mathcal{O}}_i$. So this way, our Wilson loop becomes
a matrix $W_{ij} = \langle \hat{\mathcal{O}_i} \hat{\mathcal{O}}^\dagger_j \rangle$. This can be used not only to improve the ground state overlap but also to obtain more energy levels of the system. The energy levels are the solution of the
generalized eigensystem
\begin{equation}
	\langle W_{ij}(t) \rangle c_n^j(t) = w_n \langle W_{ij}(0) \rangle c_n^j(t)\,.
	\label{varwilson}
\end{equation}

We consider again the case where the four particles form a rectangles 
and also two different alignments of the $QQ\bar{Q}\bar{Q}$
system. A parallel one, where the two quarks are on the same side of the rectangle and an antiparallel one,
where the quarks are on opposite corners of the rectangle, see Fig. \ref{geom}.

For both cases we use a basis of two operators, inspired in Ref. \cite{Bornyakov:2005kn}.
In the parallel case the operators are the tetraquark operator $\hat{\mathcal{O}}_{4Q}$, and a two meson operator. 
This gives a Wilson loop matrix where the diagonal elements are:  the tetraquark Wilson Loop
$W_{4Q} = \langle \hat{\mathcal{O}}_{4Q} (t) \hat{\mathcal{O}}_{4Q}^\dagger (0) \rangle$ (Fig. \ref{tetraq_wloop}),
the correlation of two Wilson loops (one per meson composed of a quark and an antiquark) and the off-diagonal
elements corresponding to the transition between the two states. The four matrix elements, each one corresponding to a different Wilson loop, are depicted in Fig. \ref{para_loop}.

For the antiparallel alignment, the operators we utilize are the two meson-meson operators, giving the four matrix elements
given on Fig. \ref{antip_loop}.

\section{Chromo-fields computation}

We compute the color electric and the color magnetic fields, utilizing the correlators of
the plaquettes $P_{\mu\nu}$ and the Wilson loops $W_n$.
We define the plaquettes as
$P_{\mu\nu} = 1 - \frac{1}{3} \mbox{Tr}[ U_\mu(\mathbf{s}) U_\nu(\mathbf{s}+\boldsymbol{\mu})
 U_\mu^\dagger(\mathbf{s}+\boldsymbol{\nu}) U_\nu^\dagger(\mathbf{s}) ]$.

With this definition, the chromo-fields are given by
\begin{eqnarray}
    \Braket{E^2_i} &=& \Braket{P_{0i}}-\frac{\Braket{W_n\,P_{0i}}}{\Braket{W}} \\
    \Braket{B^2_i} &=& \frac{\Braket{W\,P_{jk}}}{\Braket{W_n}}-\Braket{P_{jk}} \, ,
\end{eqnarray}
with the indices $j$ and $k$ complementing index $i$. The Lagrangian
and Energy densities are given by $\mathcal{L}=\frac{1}{2}(E^{2}-B^{2})$
and $\mathcal{H}=\frac{1}{2}(E^{2}+B^{2})$.
The Wilson loop operator for each state is given by $W_n(t) = c_n^i(t) W_{ij} c_n^j(t)$, in conformity with
Eq. (\ref{varwilson}).

The plaquettes are placed at $t = T/2$, where $T$ is the temporal extension of the Wilson Loops.
We calculate the correlators for different values of $T$ and fit the results from $T = T_{min}$ to $T = T_{max}$
to a constant, and calculate the $\chi^2/d.o.f.$.
We find that for $T_{min} = 4$ and $T_{max} = 16$, the values of the $\chi^2/d.o.f.$ are generaly acceptable, while
still giving a clear signal.
For this we consider the value of the fields as the result of the fit to a constant of the correlators
from $T = 4$ to $T = 16$.

\begin{center}
\begin{figure*}
\noindent \begin{centering}
\begin{tabular}{ccc}
\includegraphics[trim=2.1cm 0.4cm 1.1cm 0.6cm, clip, width=0.55\columnwidth]{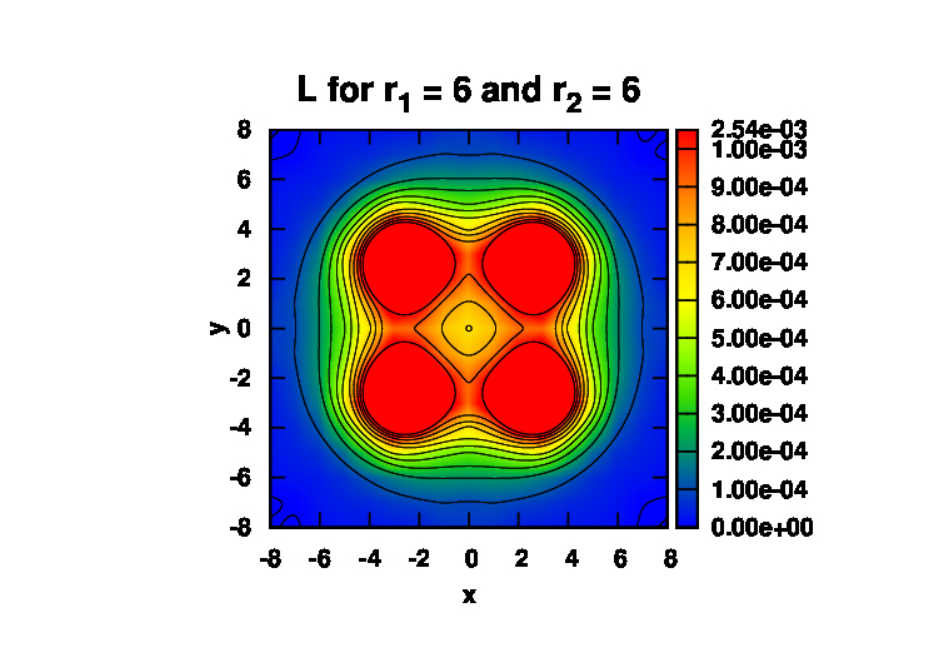} & \includegraphics[trim=2.1cm 0.4cm 1.1cm 0.6cm, clip, width=0.55\columnwidth]{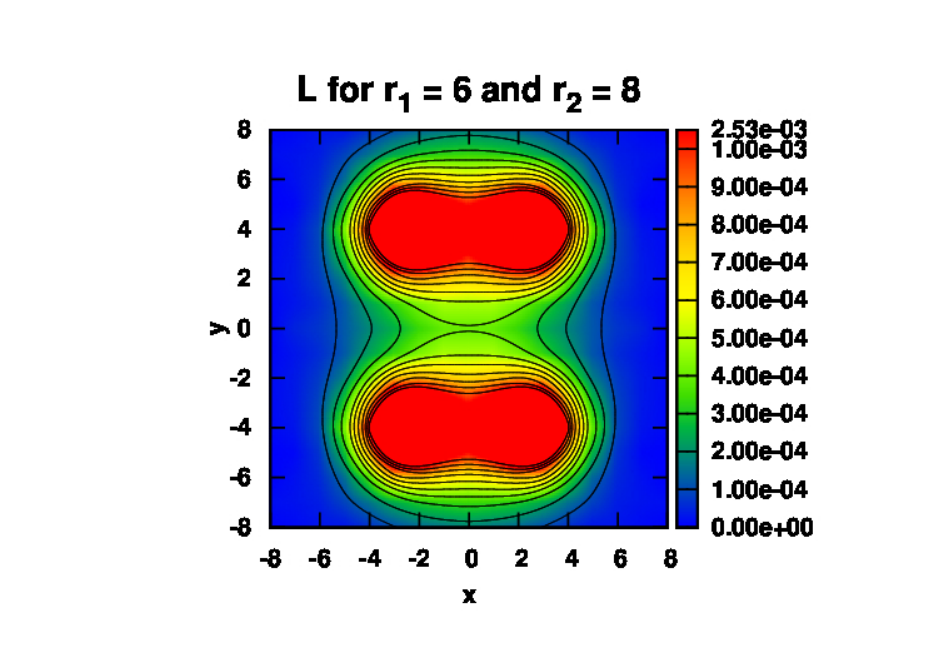} & \includegraphics[trim=2.1cm 0.4cm 1.1cm 0.6cm, clip, width=0.55\columnwidth]{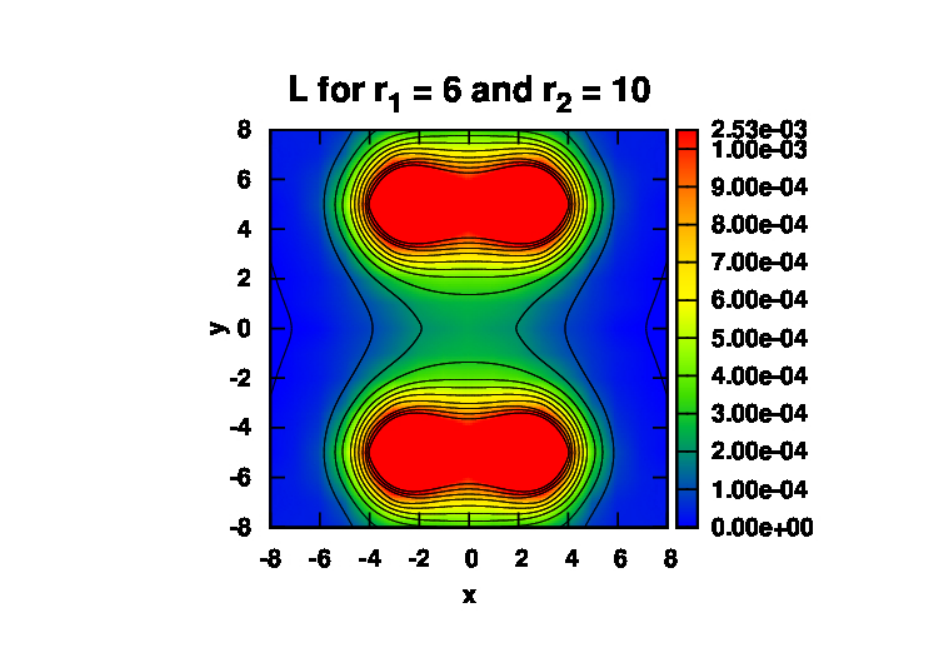}\tabularnewline
\includegraphics[trim=2.1cm 0.4cm 1.1cm 0.6cm, clip, width=0.55\columnwidth]{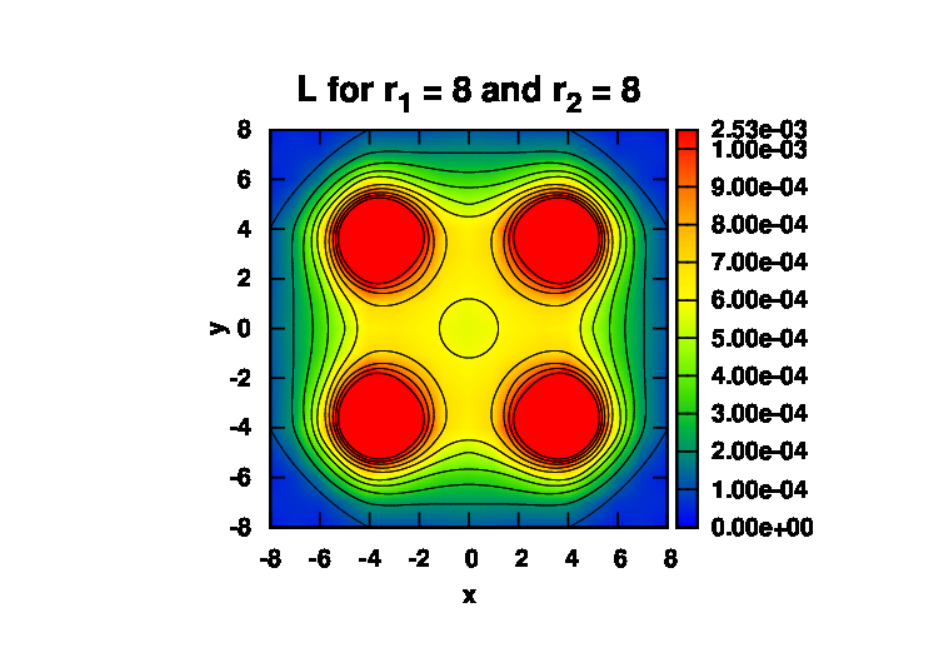} & \includegraphics[trim=2.1cm 0.4cm 1.1cm 0.6cm, clip, width=0.55\columnwidth]{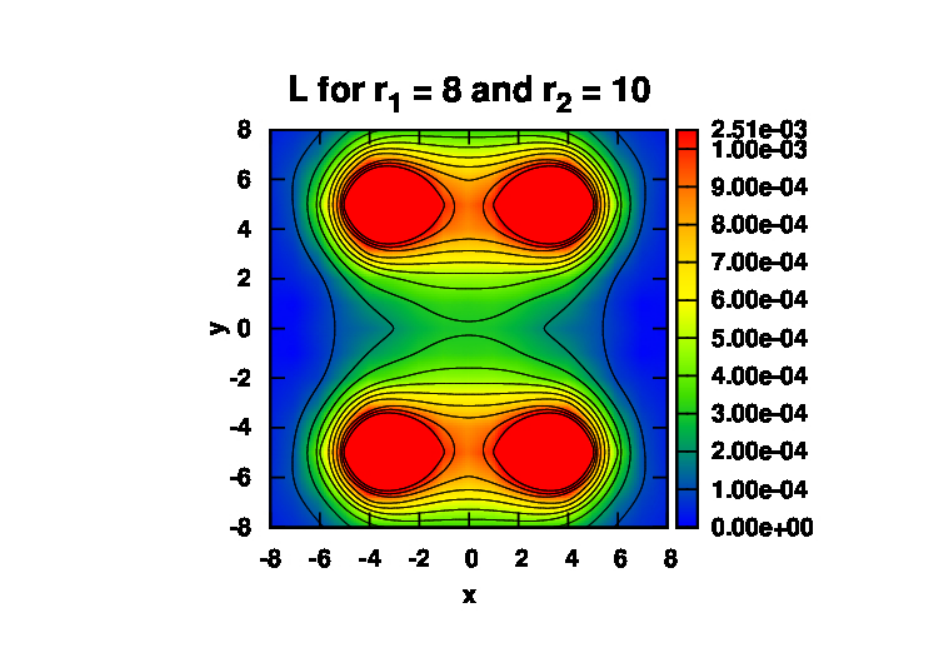} & \includegraphics[trim=2.1cm 0.4cm 1.1cm 0.6cm, clip, width=0.55\columnwidth]{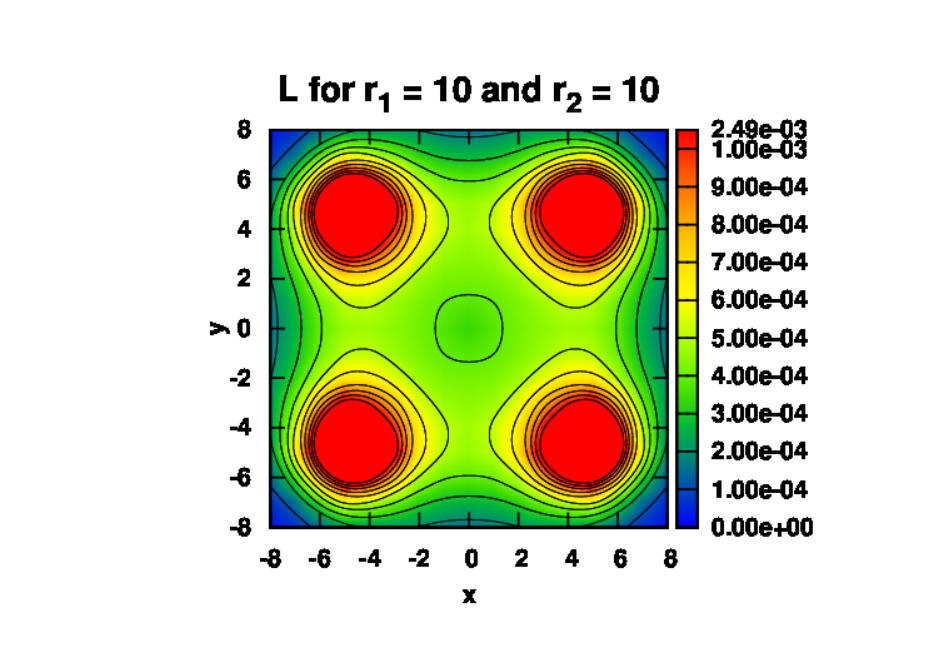}\tabularnewline
\end{tabular}
\par\end{centering}
\caption{Lagrangian density for the ground state of the antiparallel alignment.}
\label{mesonmeson_n0}%
\end{figure*}

\par\end{center}
\begin{center}
\begin{figure*}
\noindent \begin{centering}
\begin{tabular}{ccc}
\includegraphics[trim=2.1cm 0.4cm 1.1cm 0.6cm, clip, width=0.55\columnwidth]{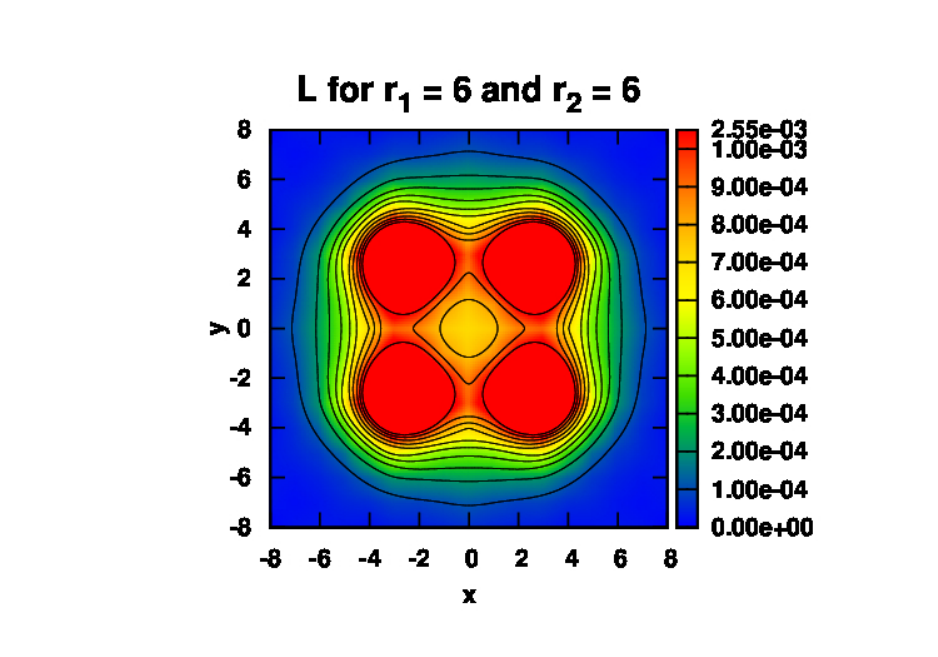} & \includegraphics[trim=2.1cm 0.4cm 1.1cm 0.6cm, clip, width=0.55\columnwidth]{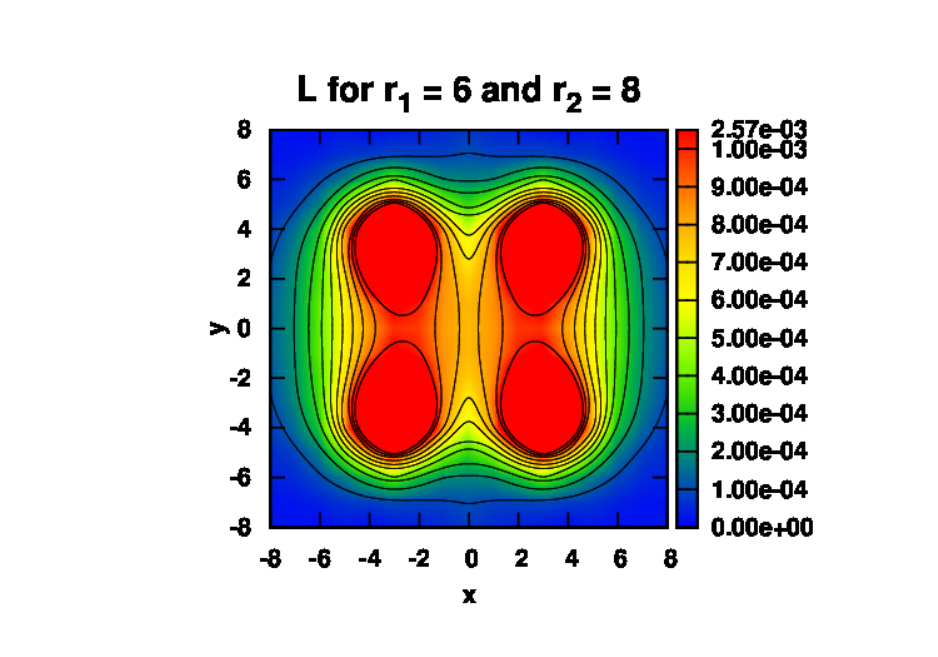} & \includegraphics[trim=2.1cm 0.4cm 1.1cm 0.6cm, clip, width=0.55\columnwidth]{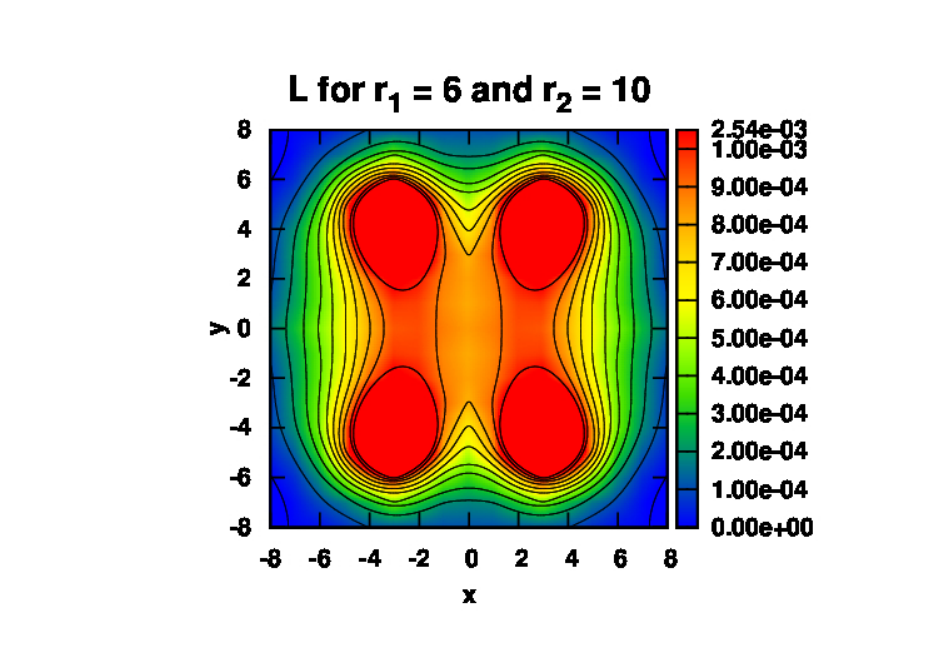}\tabularnewline
\includegraphics[trim=2.1cm 0.4cm 1.1cm 0.6cm, clip, width=0.55\columnwidth]{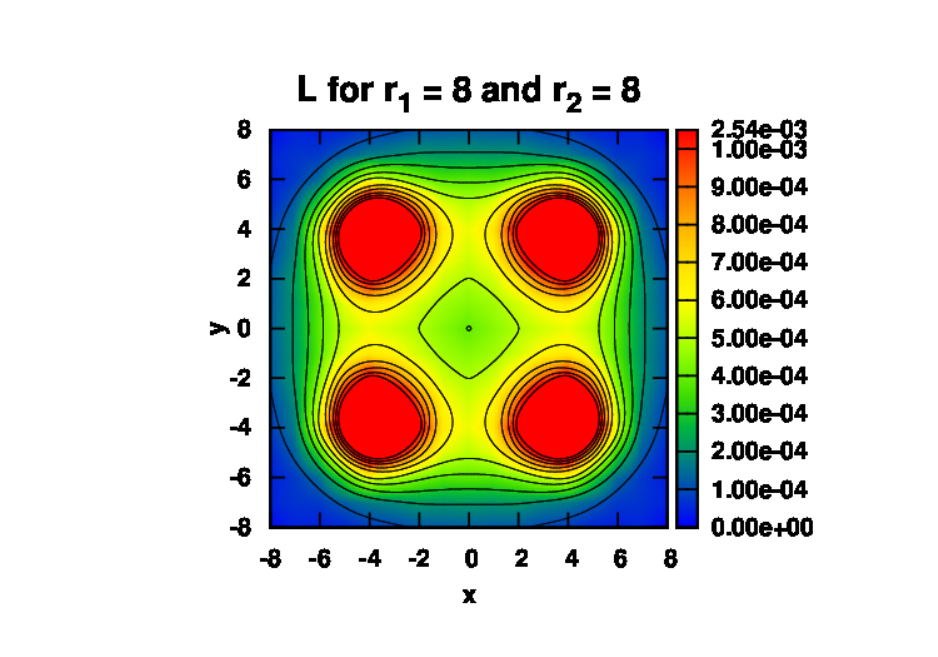} & \includegraphics[trim=2.1cm 0.4cm 1.1cm 0.6cm, clip, width=0.55\columnwidth]{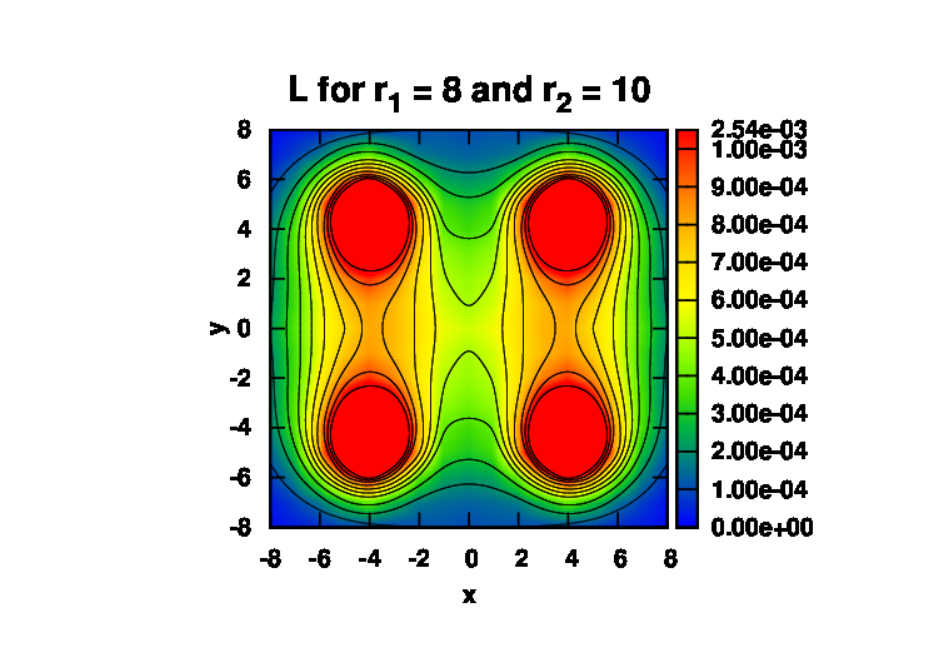} & \includegraphics[trim=2.1cm 0.4cm 1.1cm 0.6cm, clip, width=0.55\columnwidth]{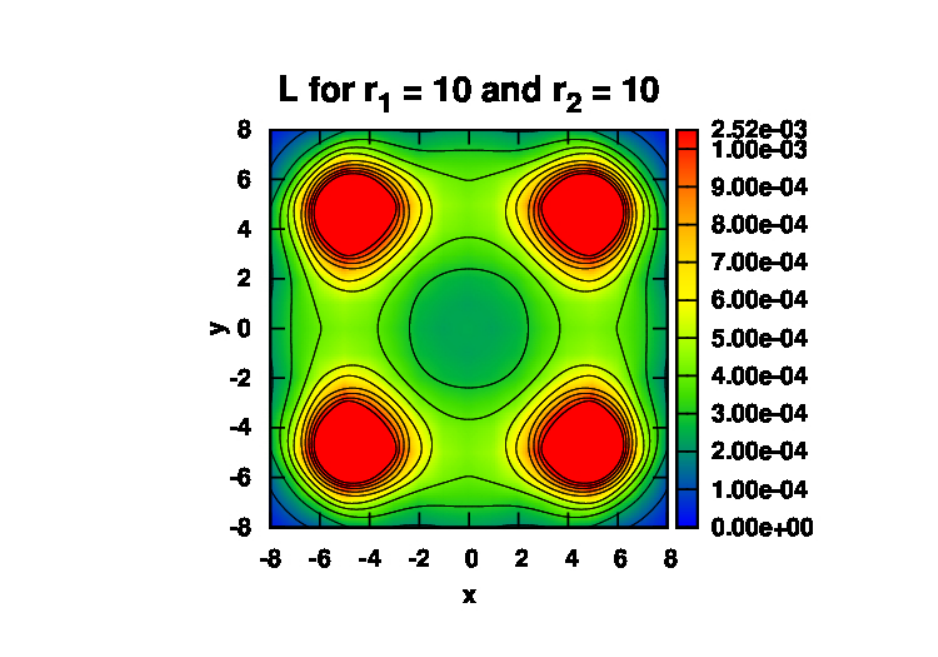}\tabularnewline
\end{tabular}
\par\end{centering}
\caption{Lagrangian density for the first excited state of the antiparallel
alignment.}
\label{mesonmeson_n1}%
\end{figure*}
\par\end{center}

\section{Results for the Color fields}

In this work, we only consider the geometry of the four particles in a plane forming a rectangle with two distinct kinds of alignment, see Fig. \ref{geom}. 
When the two quarks/antiquarks are on the same side of the rectangle, this corresponds to a parallel alignment; if they are in opposite sides, we have an antiparallel alignment. 
The parallel alignment is useful in the visualization of the transition between the tetraquark and the two meson states,
while the antiparallel one is be used to observe the transition between the two meson-meson states.

\begin{figure*}
\begin{tabular}{cccc}
\includegraphics[trim=2.1cm 0.4cm 1.1cm 0.6cm, clip,width=0.5\columnwidth]{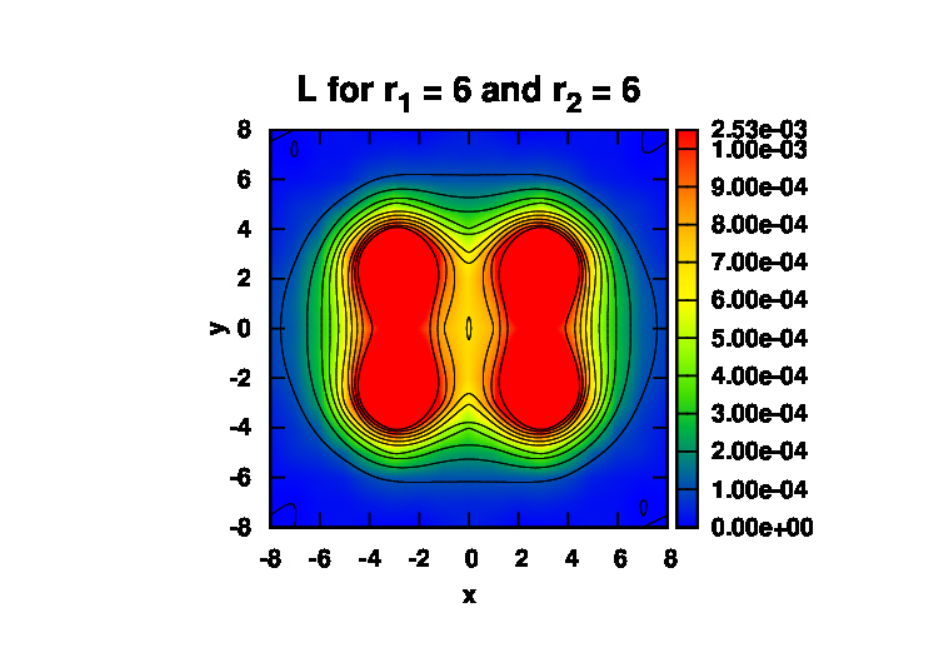} & \includegraphics[trim=2.1cm 0.4cm 1.1cm 0.6cm, clip,width=0.5\columnwidth]{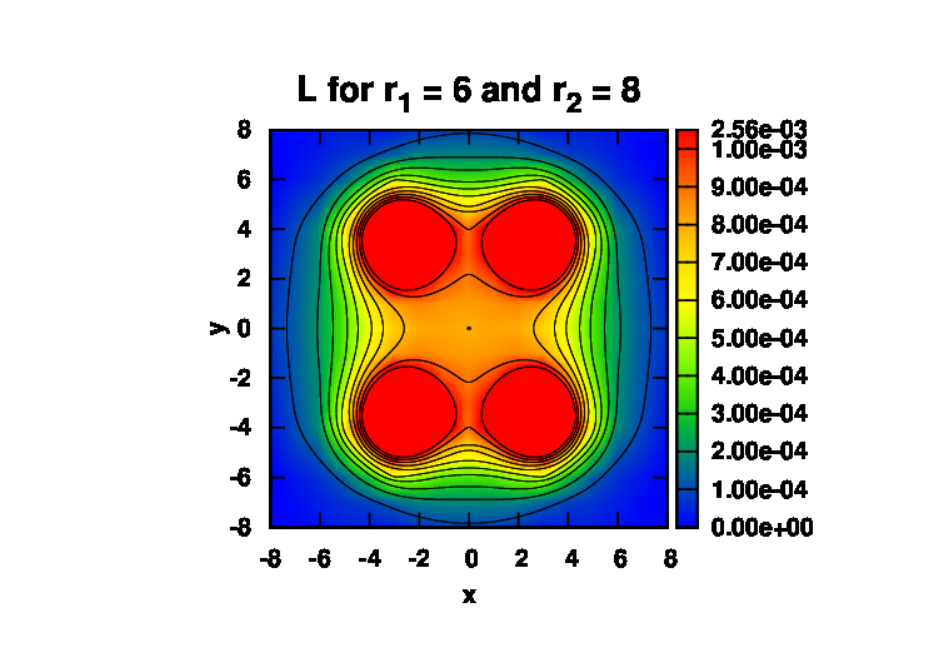} & \includegraphics[trim=2.1cm 0.4cm 1.1cm 0.6cm, clip,width=0.5\columnwidth]{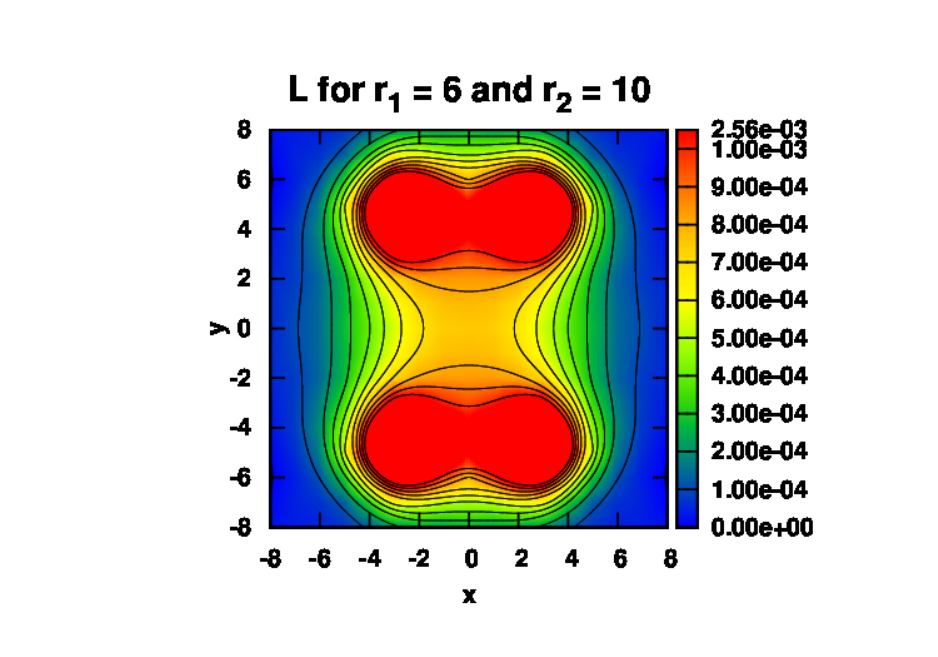} & \includegraphics[trim=2.1cm 0.4cm 1.1cm 0.6cm, clip,width=0.5\columnwidth]{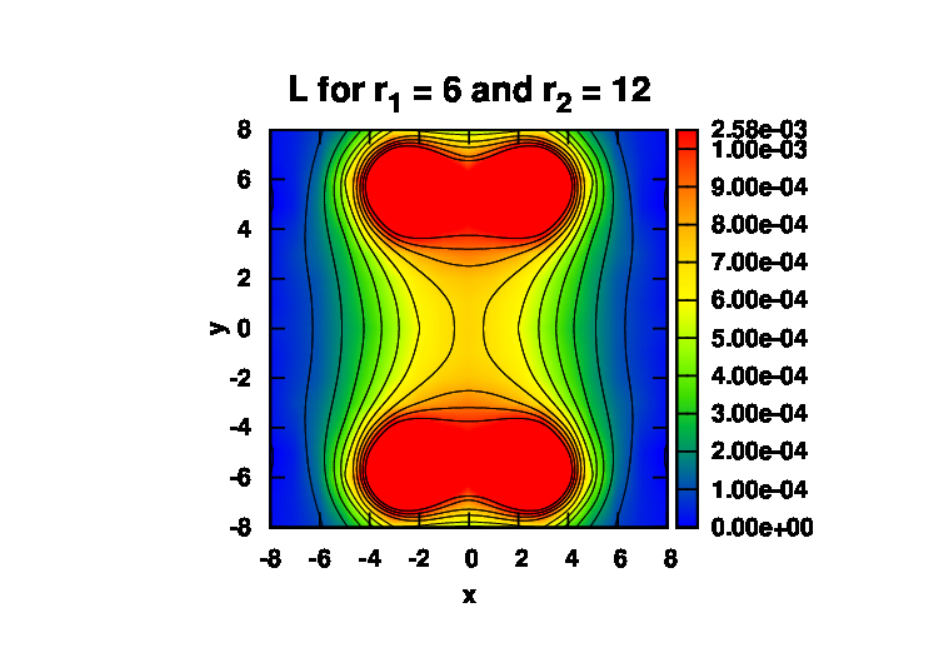}\tabularnewline
\includegraphics[trim=2.1cm 0.4cm 1.1cm 0.6cm, clip,width=0.5\columnwidth]{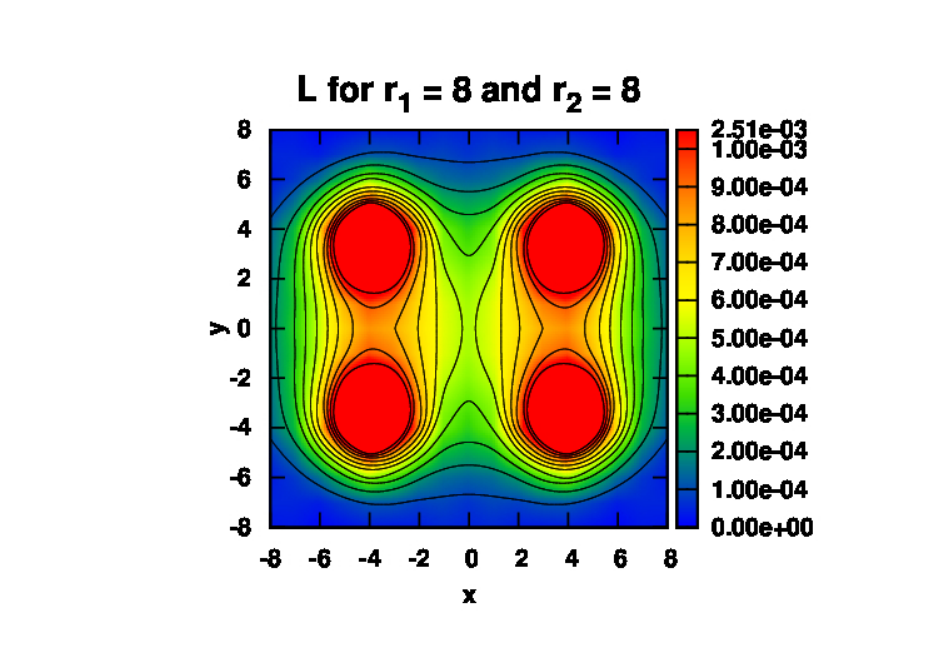} & \includegraphics[trim=2.1cm 0.4cm 1.1cm 0.6cm, clip,width=0.5\columnwidth]{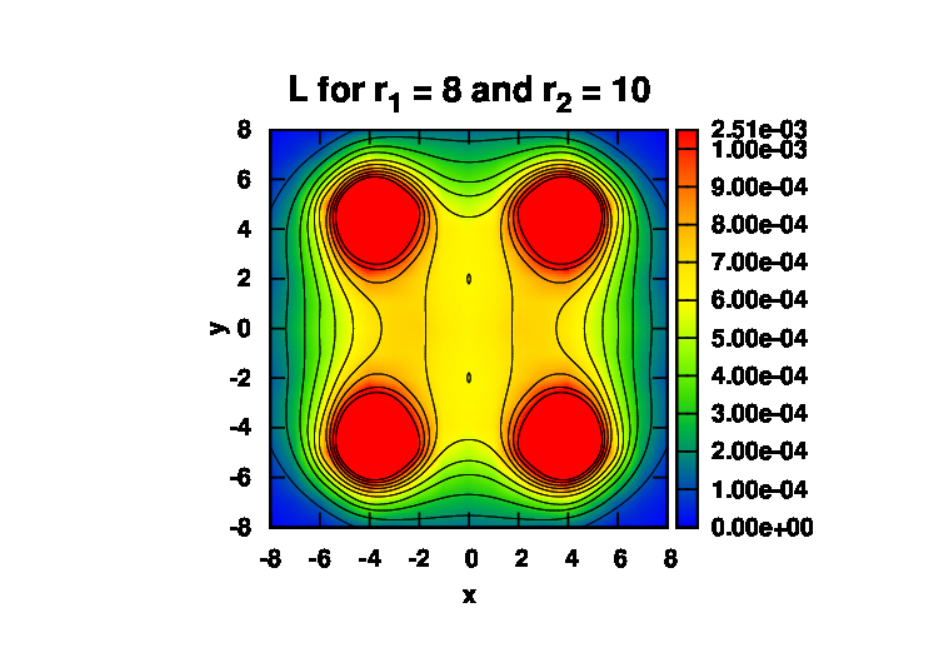} & \includegraphics[trim=2.1cm 0.4cm 1.1cm 0.6cm, clip,width=0.5\columnwidth]{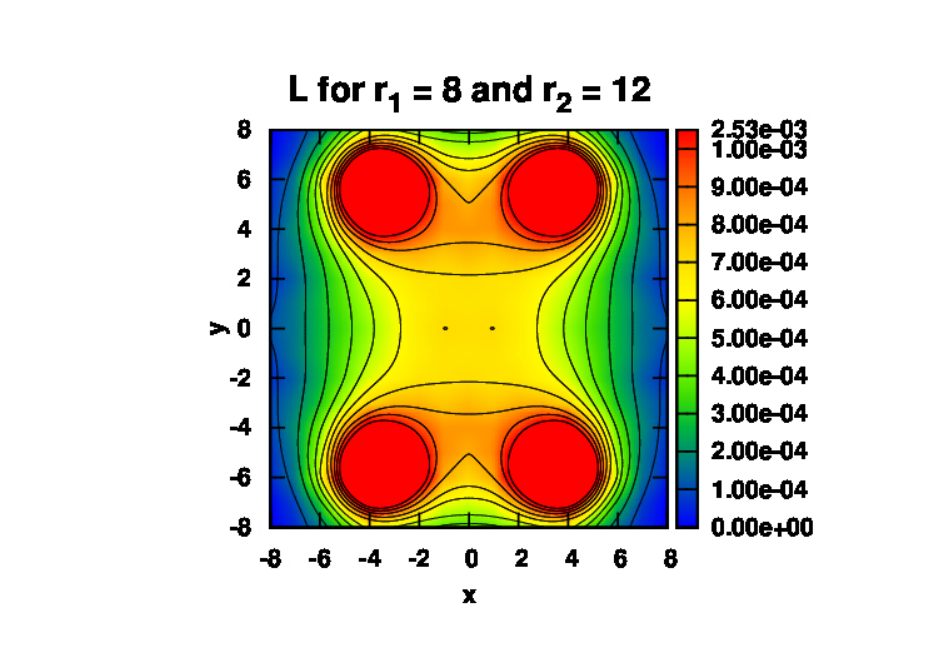} & \includegraphics[trim=2.1cm 0.4cm 1.1cm 0.6cm, clip,width=0.5\columnwidth]{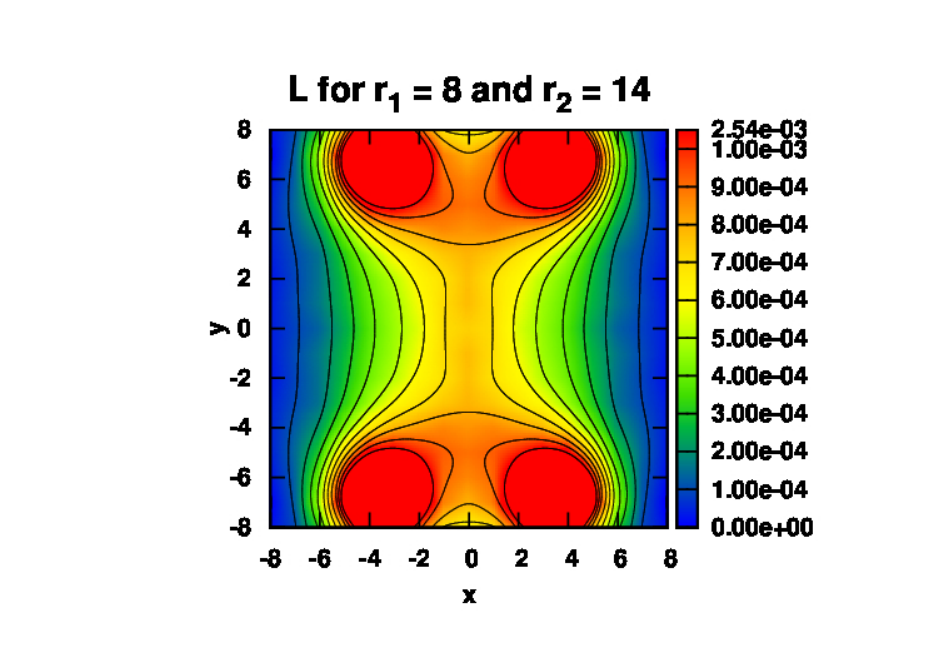}\tabularnewline
\end{tabular}
\caption{Lagrangian density for the ground state of the parallel alignment.}
\label{tetraq_n0}
\end{figure*}

\begin{figure*}
\begin{tabular}{cccc}
\includegraphics[trim=2.1cm 0.4cm 1.1cm 0.6cm, clip,width=0.5\columnwidth]{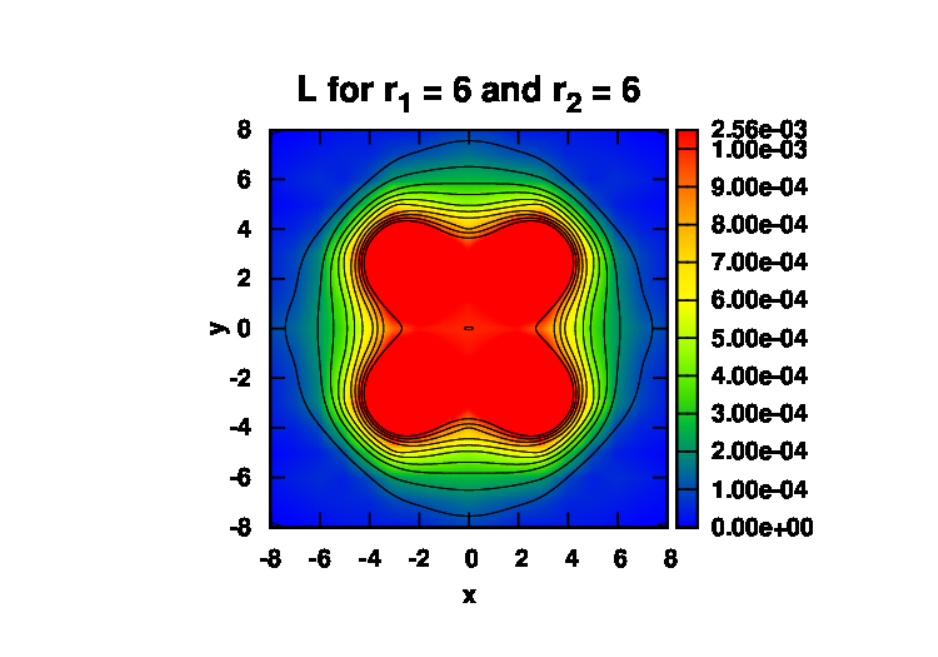} & \includegraphics[trim=2.1cm 0.4cm 1.1cm 0.6cm, clip,width=0.5\columnwidth]{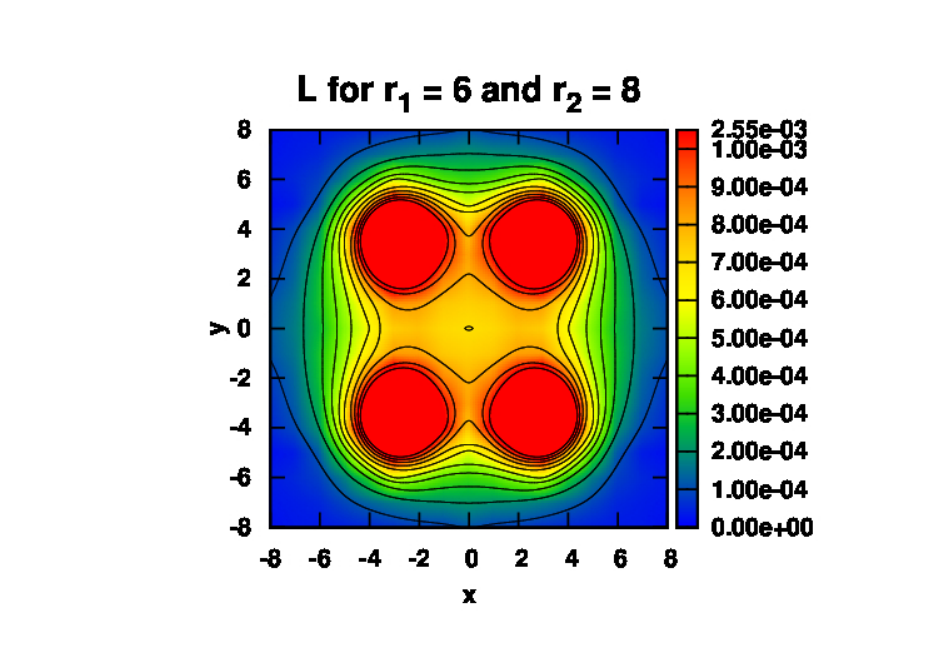} & \includegraphics[trim=2.1cm 0.4cm 1.1cm 0.6cm, clip,width=0.5\columnwidth]{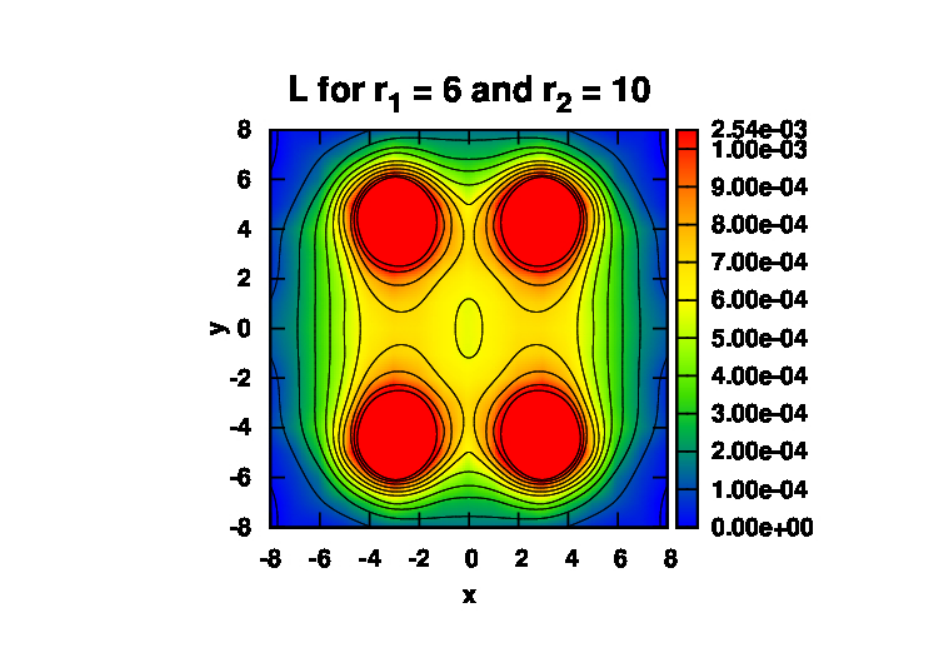} & \includegraphics[trim=2.1cm 0.4cm 1.1cm 0.6cm, clip,width=0.5\columnwidth]{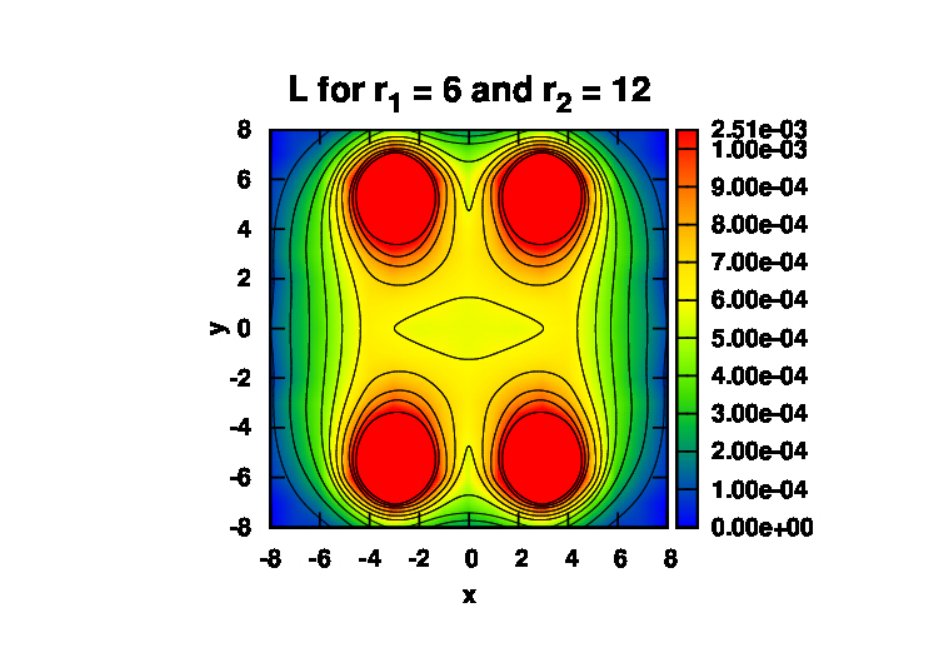}\tabularnewline
\includegraphics[trim=2.1cm 0.4cm 1.1cm 0.6cm, clip,width=0.5\columnwidth]{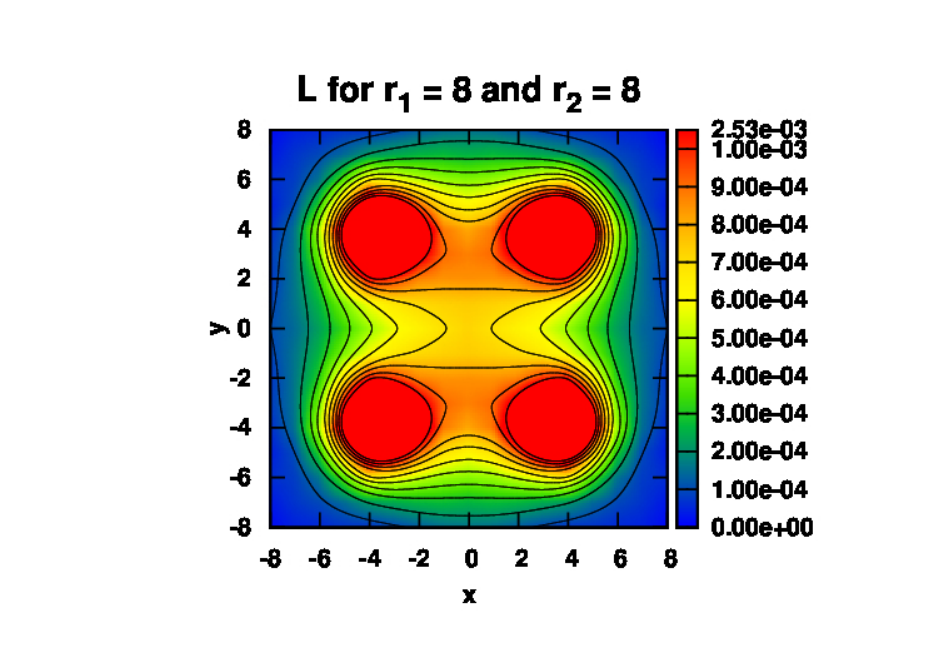} & \includegraphics[trim=2.1cm 0.4cm 1.1cm 0.6cm, clip,width=0.5\columnwidth]{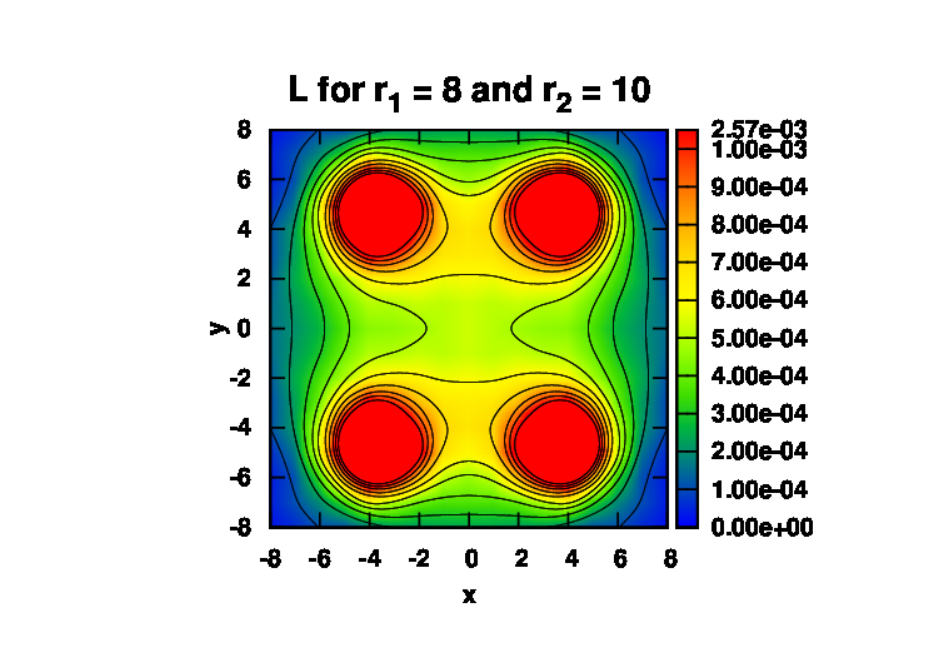} & \includegraphics[trim=2.1cm 0.4cm 1.1cm 0.6cm, clip,width=0.5\columnwidth]{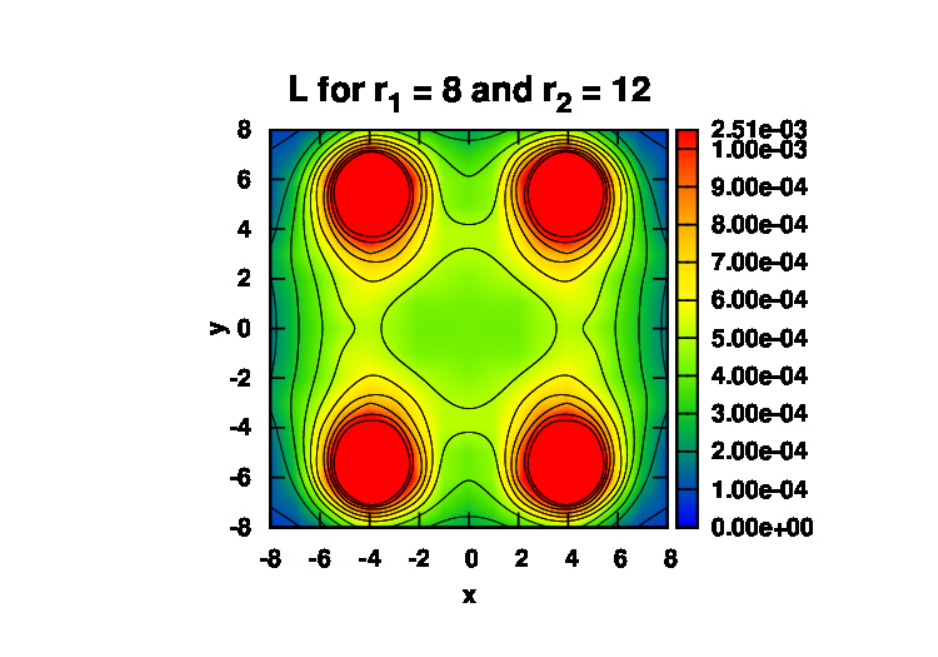} & \includegraphics[trim=2.1cm 0.4cm 1.1cm 0.6cm, clip,width=0.5\columnwidth]{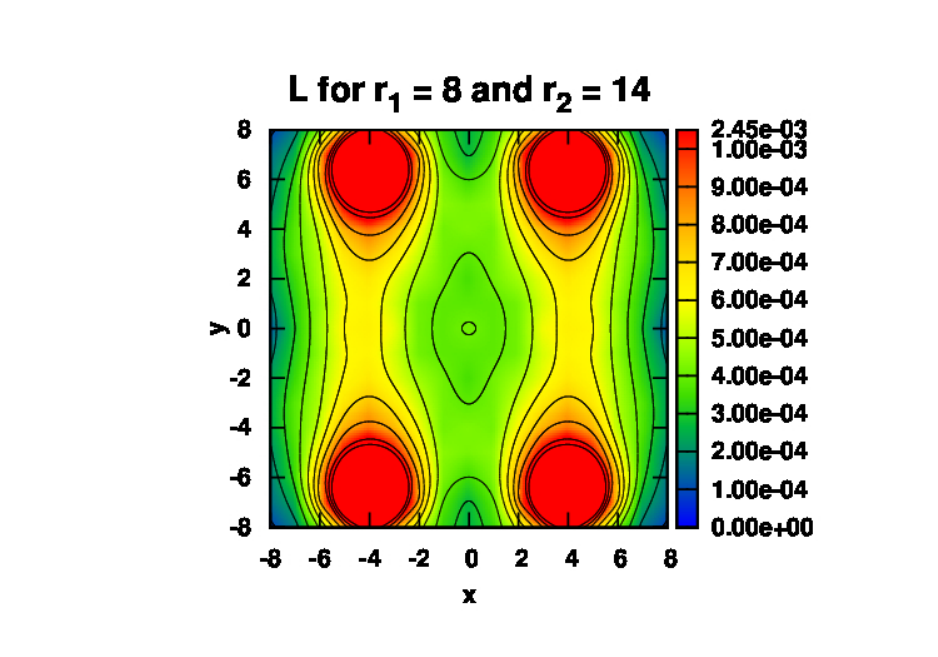}\tabularnewline
\end{tabular}
\caption{Lagrangian density for the first excited state of the parallel alignment.}
\label{tetraq_n1}
\end{figure*}

The results presented in this work are obtained using 1121 quenched lattice QCD configurations with dimension $24^3 \times 48$ and $\beta = 6.2$ with lattice spacing, $a$, $a = 0.07261(85)$ fm or $a^{-1} = 2718(32)$ MeV
The configurations were generated with GPUs using a combination of Cabbibo-Marinari, pseudo-heatbath and over-relaxation algorithms, \cite{Cardoso:2011xu,ptqcd}.
APE smearing \cite{Cardoso:2009kz} and Hypercubic blocking \cite{Hasenfratz:2001hp} are used to improve the signal to noise ratio.

\begin{figure*}
\begin{tabular}{cc}
\includegraphics[width=0.95\columnwidth]{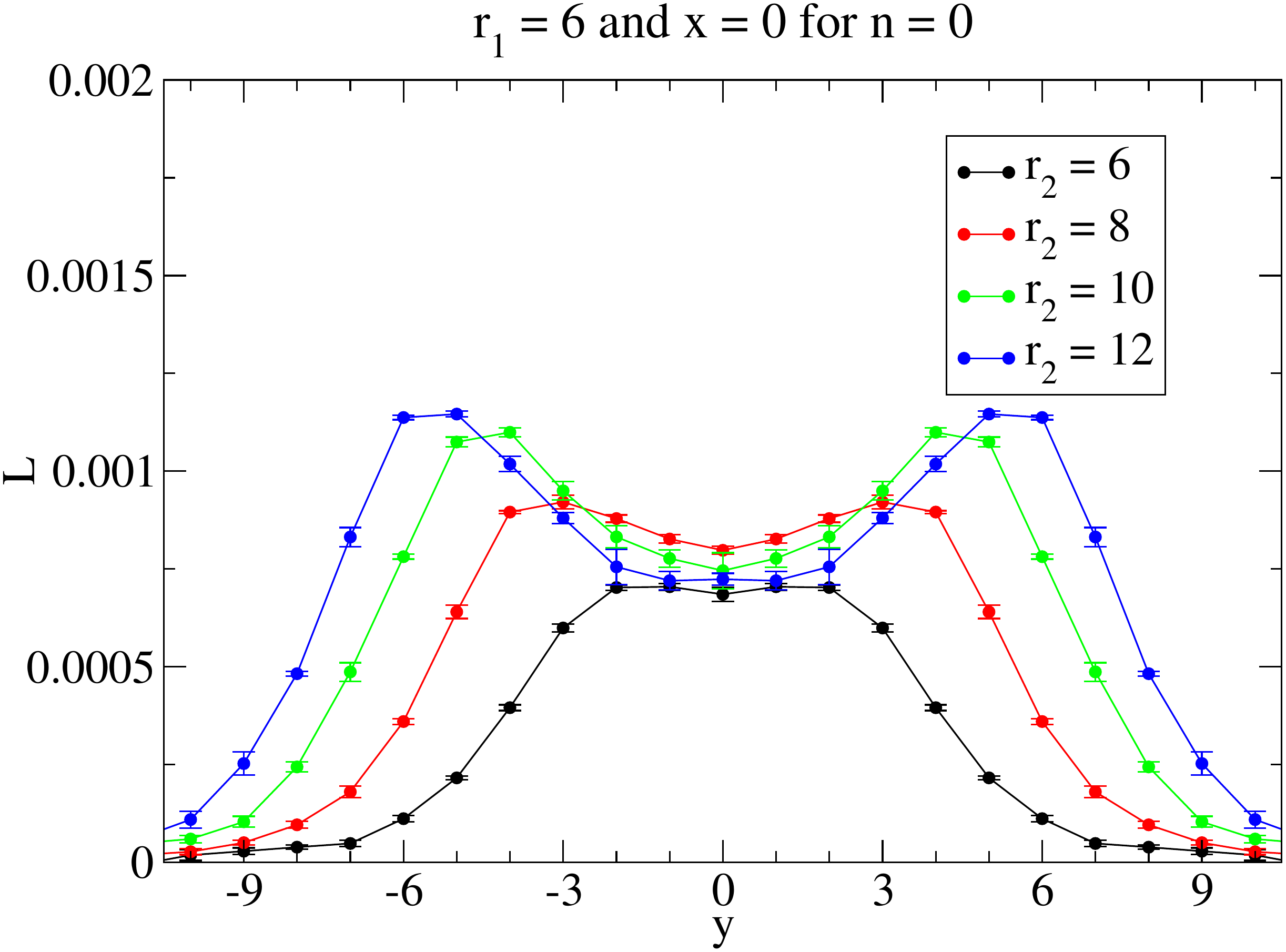}\label{tetraq6_cuts:a} & \includegraphics[width=0.95\columnwidth]{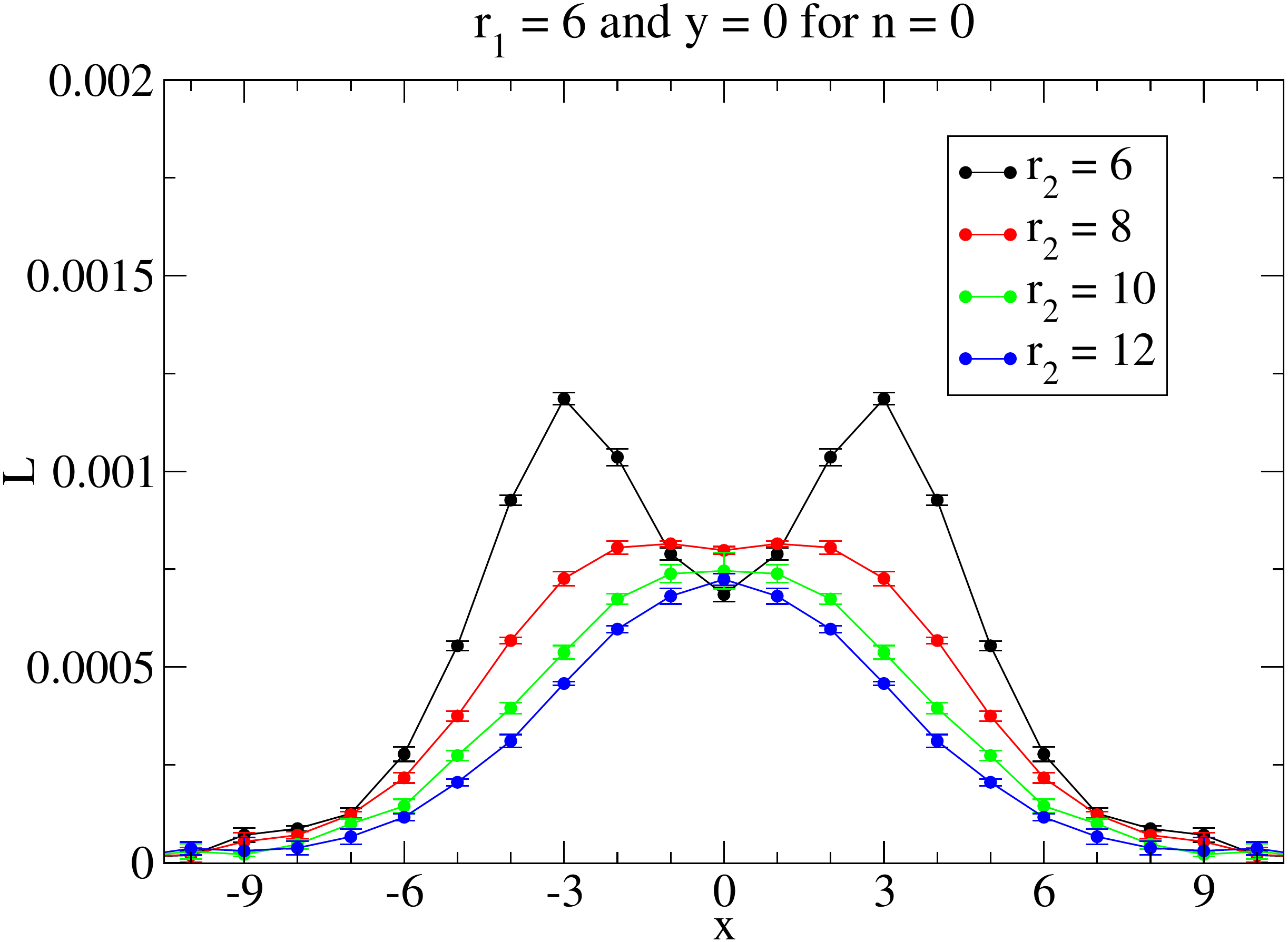}\label{tetraq6_cuts:b}\tabularnewline
\includegraphics[width=0.95\columnwidth]{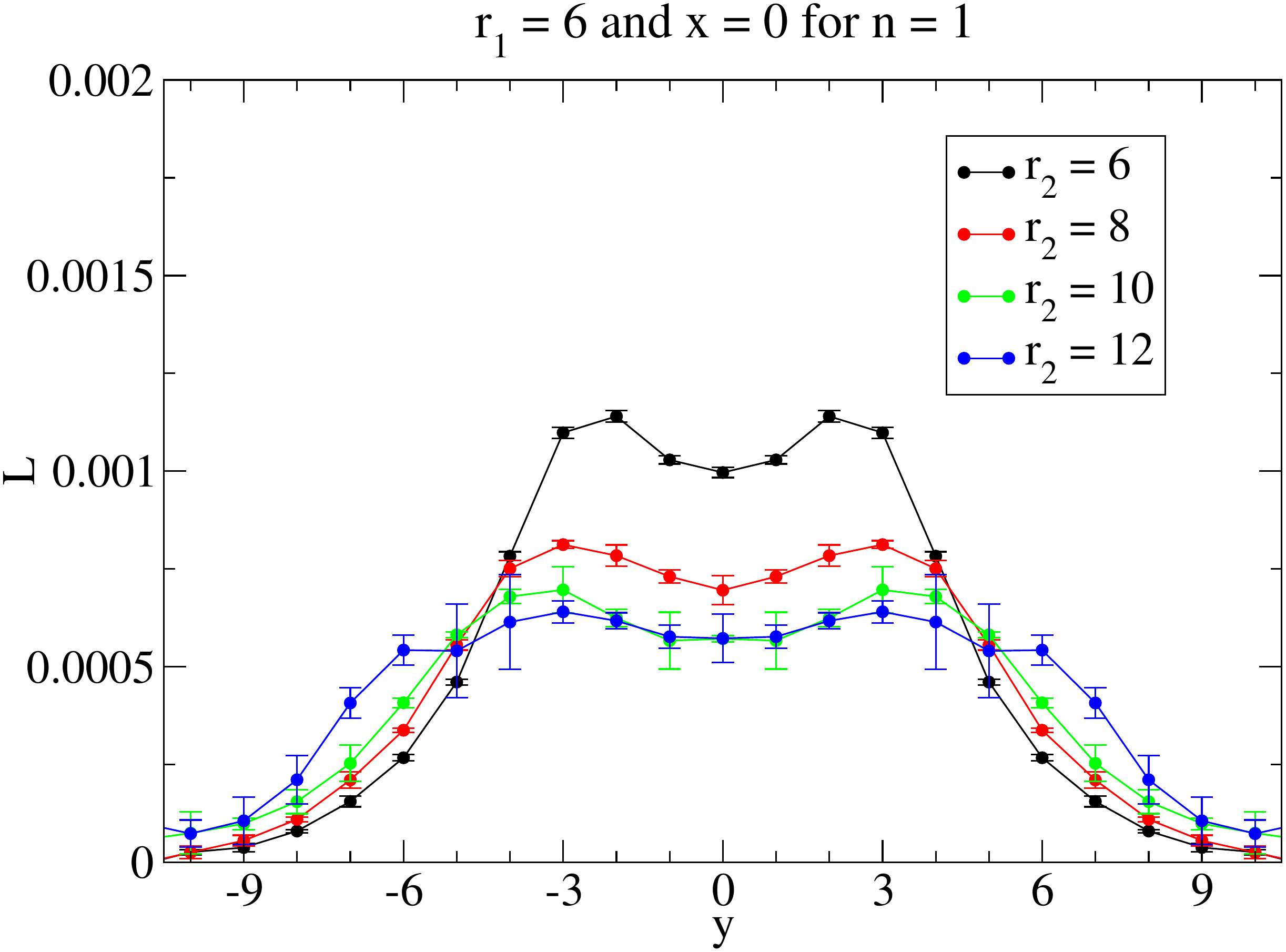}\label{tetraq6_cuts:c} & \includegraphics[width=0.95\columnwidth]{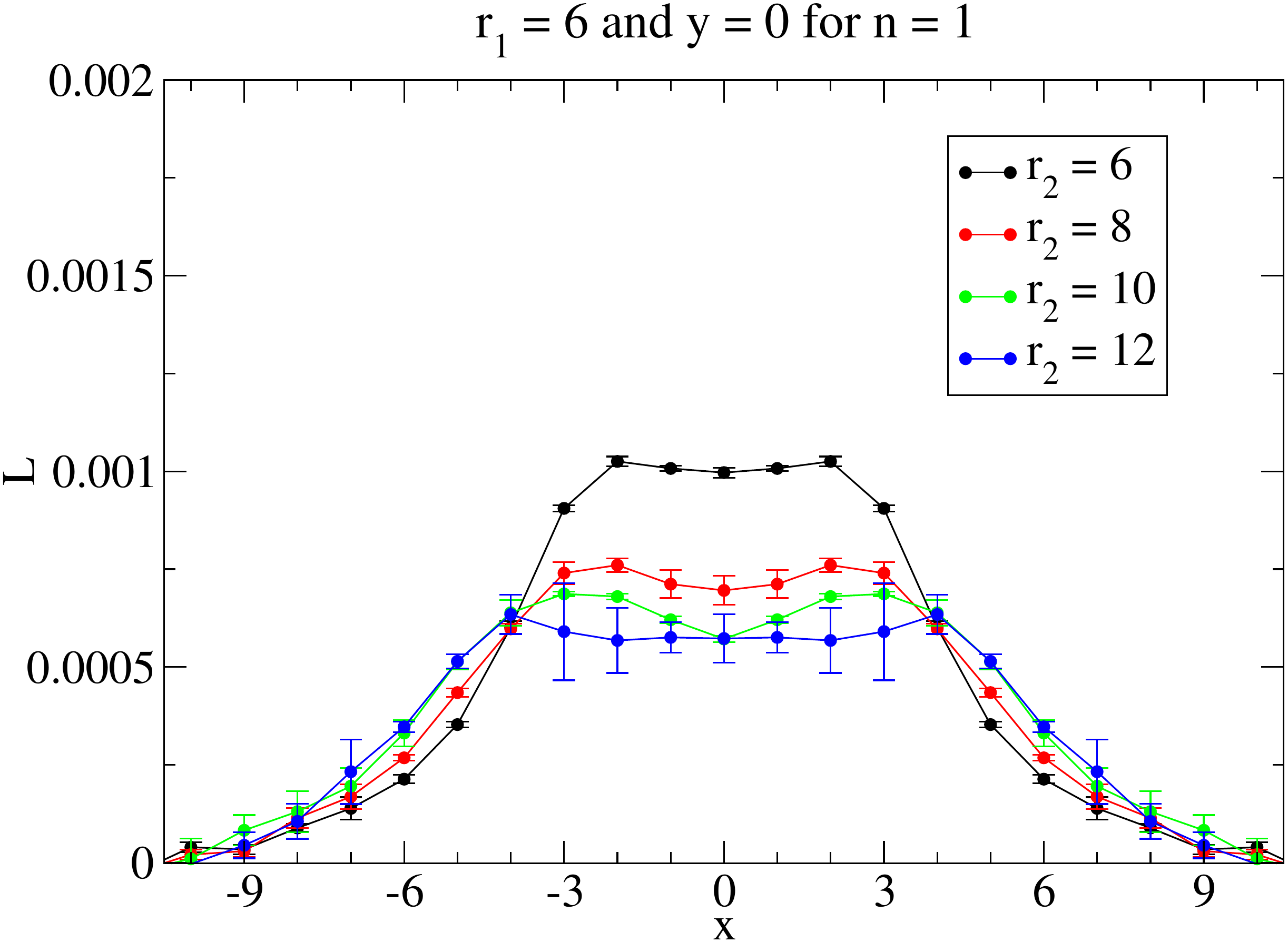}\label{tetraq6_cuts:d}\tabularnewline
\end{tabular}
\caption{Cuts of the Lagrangian density for $x = 0$ and for $y = 0$, with $r_{1} = 6$,
both for the ground state and for the first excited state.
\label{tetraq6_cuts} }
\end{figure*}

\begin{figure*}
\begin{tabular}{cc}
\includegraphics[width=0.95\columnwidth]{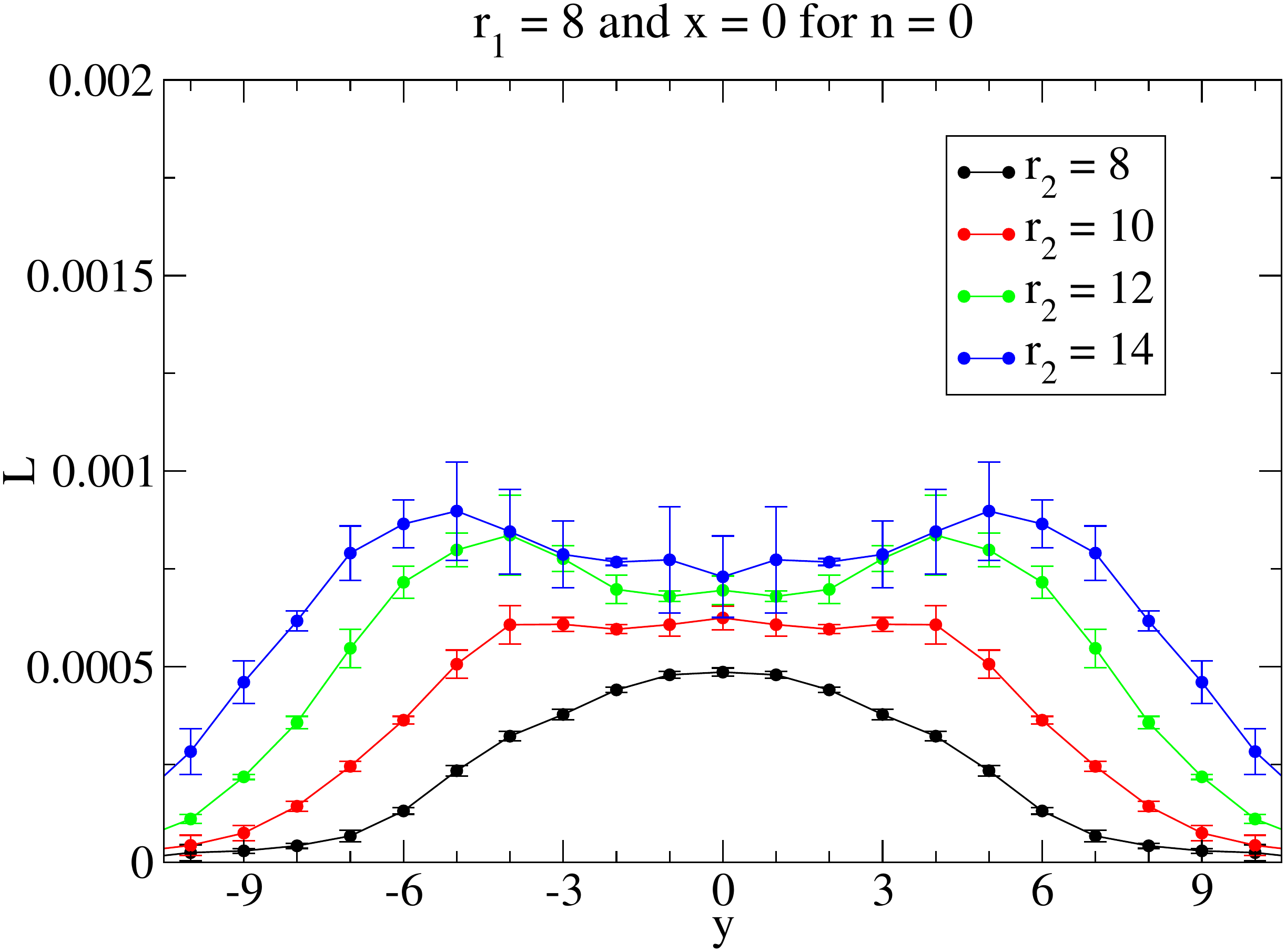}\label{tetraq8_cuts:a} & \includegraphics[width=0.95\columnwidth]{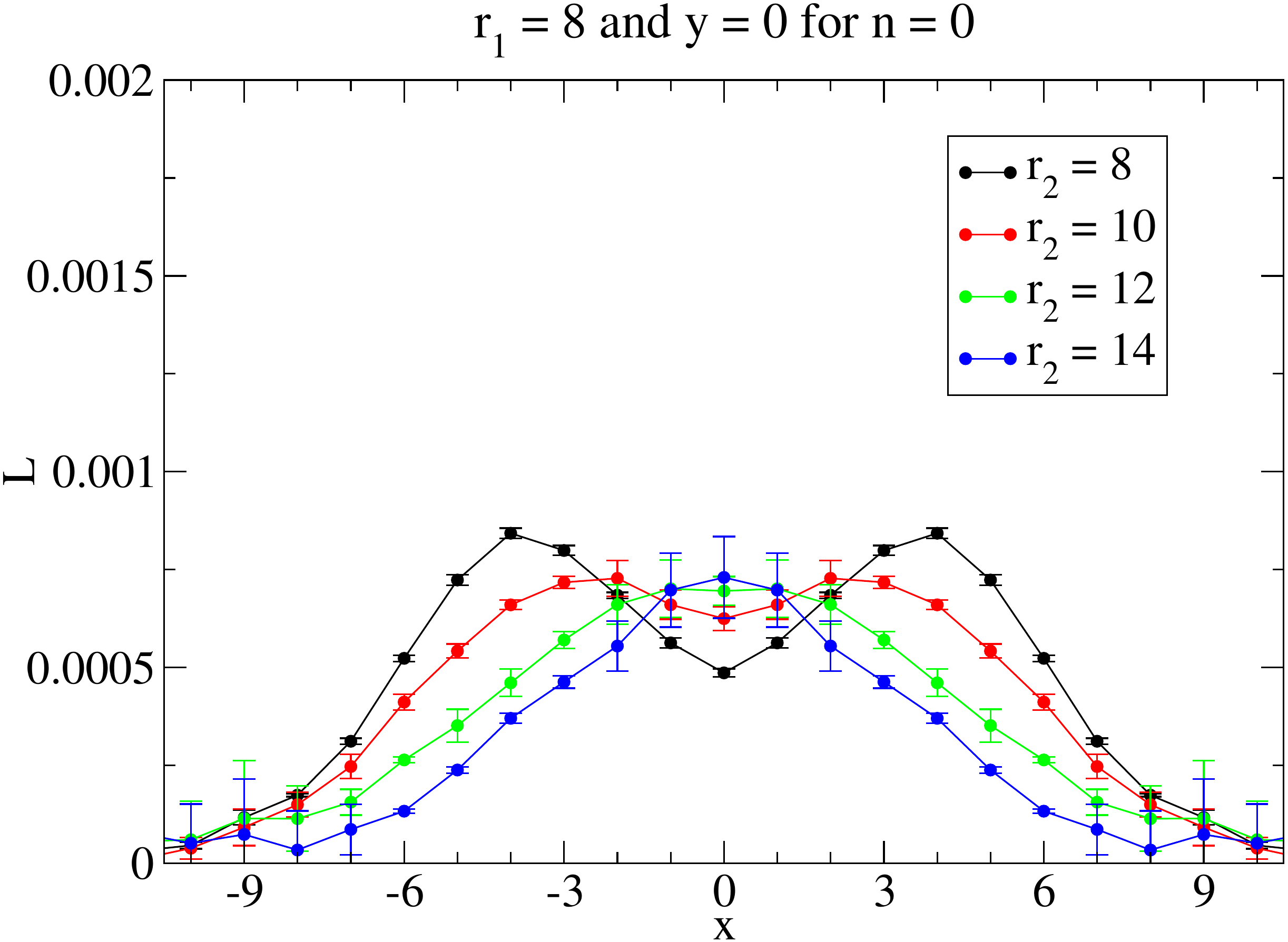}\label{tetraq8_cuts:b}\tabularnewline
\includegraphics[width=0.95\columnwidth]{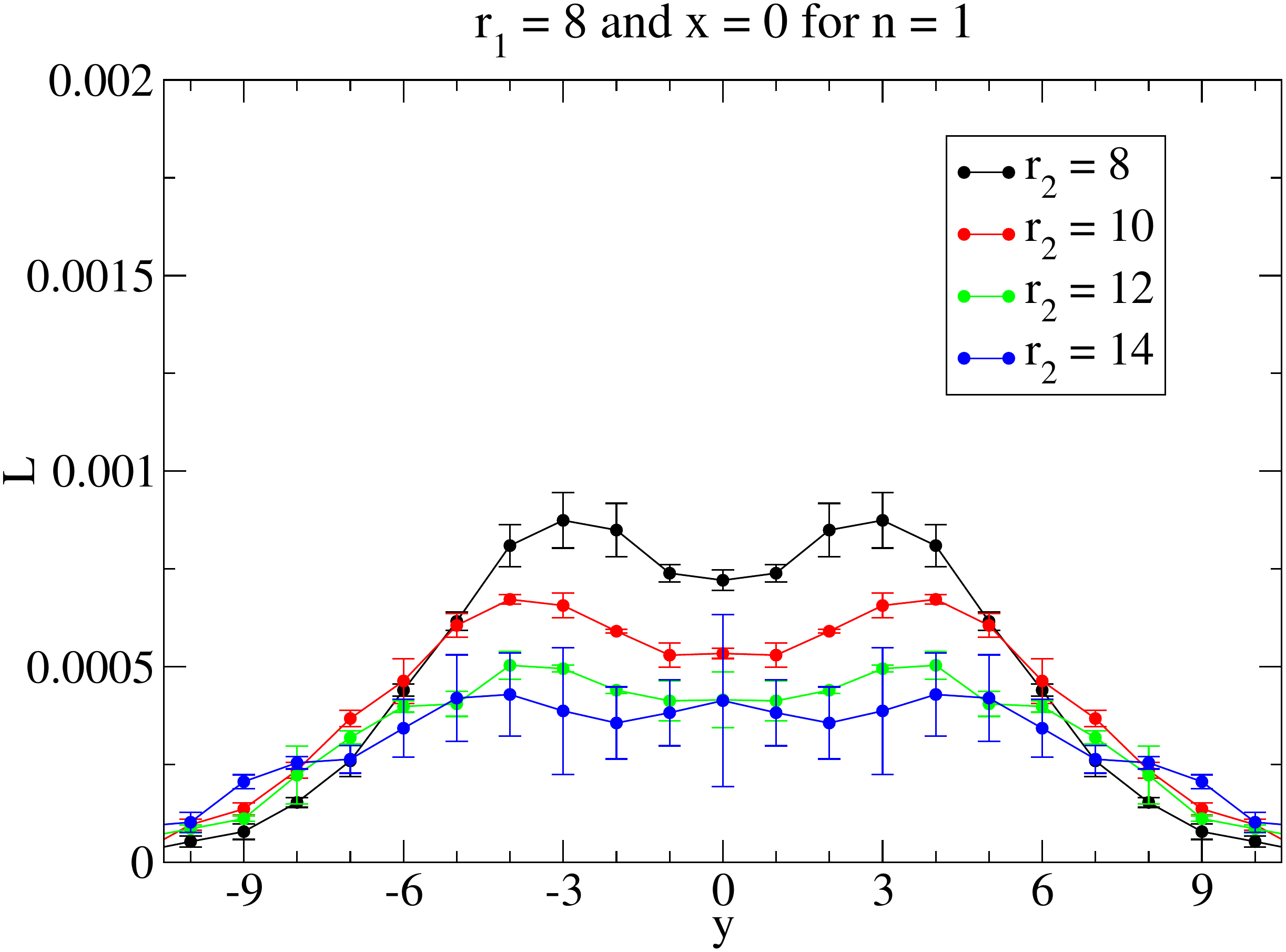}\label{tetraq8_cuts:c} & \includegraphics[width=0.95\columnwidth]{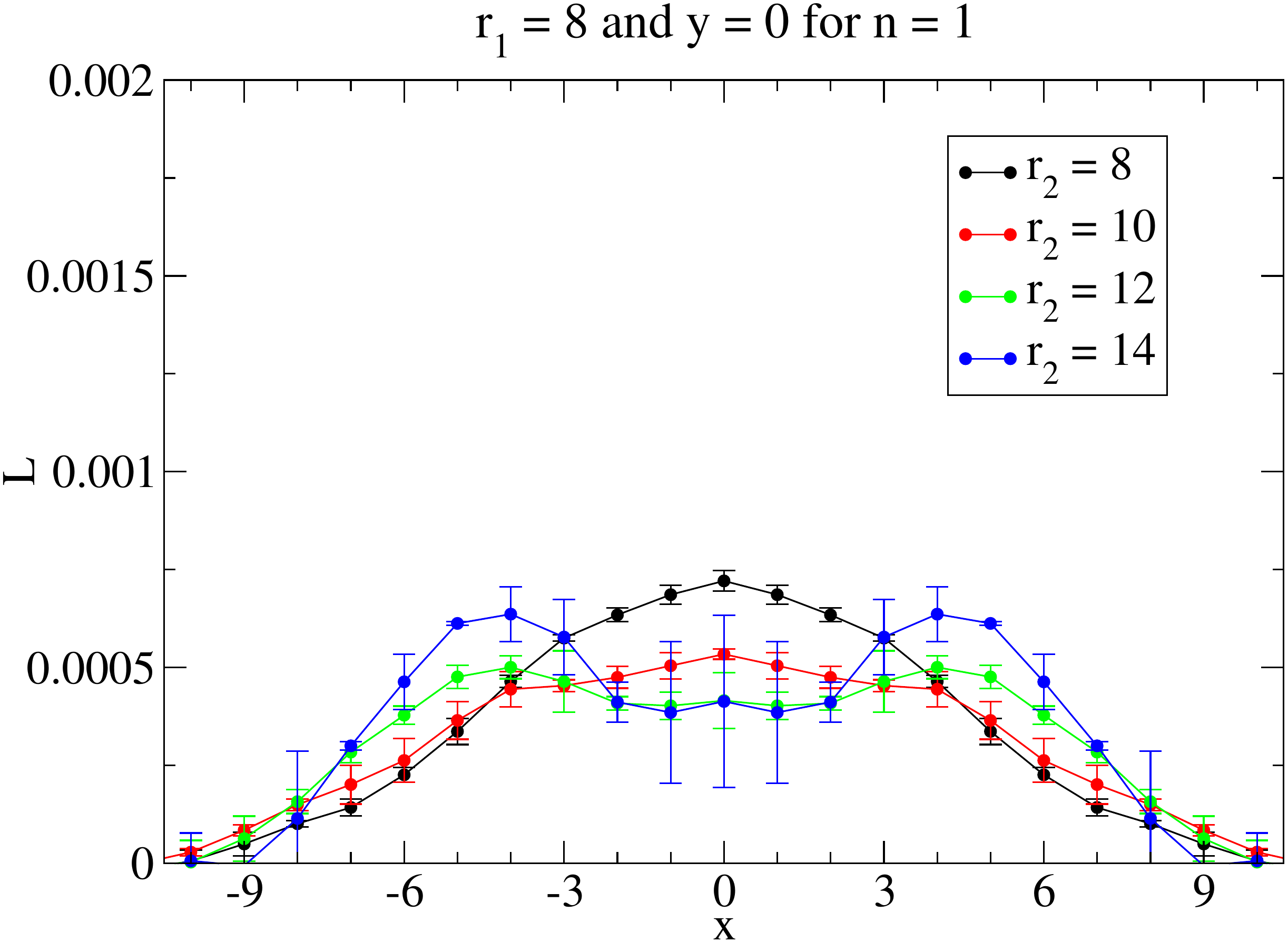}\label{tetraq8_cuts:d}\tabularnewline
\end{tabular}
\caption{Cuts of the Lagrangian density for $x = 0$ and for $y = 0$, for $r_{1}=8$,
both for the ground state and for the first excited state.
\label{tetraq8_cuts}}
\end{figure*}

In Figs. \ref{mesonmeson_n0} and \ref{mesonmeson_n1}, we show the results of the Lagrangian density for the antiparallel alignment for the ground state, $n=0$, and the first excited state, $n=1$, respectively, with different values of $r_{1}$ and $r_{2}$.
Note that, in this geometry the system is symmetric for the exchange of $r_{1}$ and $r_{2}$. 
When $r_{1}=r_{2}$, the system forms a square-symmetric structure for both the ground and first excited states, thus the two states have to correspond to the symmetric and antisymmetric color wavefunctions.
Inspecting the composition of the variational Wilson Loop for the two states, we are able to conclude that the ground
state corresponds in this geometry to the color symmetric wavefunction, with the color antisymmetric corresponding to the first excited state.
This is the  reverse of what happens in the tetraquark sector, where the ground state is color antisymmetric.
When $r_{1} < r_{2}$, the groundstate results show a pure two-meson ground state.
In what concerns the first excited state, it must have a color wavefunction  orthogonal to the one of the ground state. We observe a first excited state different from the other two-meson state, where the confining string seems to be linking all the four particles together.

In Figs. \ref{tetraq_n0} and \ref{tetraq_n1}, we depict the plot of the Lagrangian
density of the parallel alignment, in the ground and first excited states respectively.
These results are complemented with the Figs. \ref{tetraq6_cuts} and \ref{tetraq8_cuts}, with cuts cuts for $x = 0$ and $y = 0$, for $r_1 = 6$ and $r_2 = 8$.

We now analyse the ground states profiles.
In upper graphics of both Fig. \ref{tetraq6_cuts} and \ref{tetraq8_cuts},
we see the results for the cuts of the Lagrangian density (with $r_1 = 6$ and $r_1 = 8$ respectively),
for $x = 0$ (on the left side) and $y = 0$ (on the right side).
The first case corresponds to the center of the predictable diquark-diantiquark
flux tube formed in the tetraquark domain, while the second
corresponds to a transversal cut of the same flux tube and also of the two mesonic
flux tubes, whose the centers should be in $-r_1 / 2 $ and $r_2 / 2$.
We observe the transition between the tetraquark and the two meson states.

More interesting and not so easily interpretable is the first excited state.
We can with a fair amount of certainty claim that this first excited state is due to the
recombination of the flux tubes and not due to the flux tube excitations as in
 \cite{luscher-2004-0407,Juge:2002br}.
For the distances studied, we always seem to have a fully confined system, with all the four particles connected by the flux tube (Figs. \ref{mesonmeson_n1}, \ref{tetraq_n1}, \ref{tetraq6_cuts} and \ref{tetraq8_cuts}) contrarily to what we can naively be expect for the meson to meson transition.
The first excited state clearly is not the other two meson state, but a different color state.
Neither, is it a vibrational excitation of a flux tube.

Indeed the first excited state should be orthogonal in the color space to the ground state.
Since both the ground state and the first excited state are eigenvectors of the static potential matrix since the phenomenon which is happening is the recombination of the flux tube, 
and since this matrix is hermitian, the two vectors have to be orthogonal to each other.
Consequently, we expect that $|\bar{I}\rangle$ to be the color vector of the first excited
state, when the ground state has the color of the meson system $|I\rangle$.
Similarly, we expect that the first excited state to be given by the symmetric color state
$|S\rangle$, when the ground state is the antisymmetric one $|A\rangle$. This happens when we are in the tetraquark domain.

\section{Discussion}

As we discussed before, the results seem to agree with the previous ones, obtained for the static potential \cite{Alexandrou:2004ak,Okiharu:2004ve}, which support the generalized flip-flop picture for the ground state of two quarks and two antiquarks system.
Namely, we observe the formation of the tetraquark string and of the
two mesonic strings in the domains we would expect that to happen.

For the first excited states, the situation is not as clear, and so to better understand these states
we calculate the Casimir scaling factors for the different color wavefunctions in Table \ref{casimir}.
This shall provide us with qualitative insight to what kind of interaction we expect in the system.

In Table \ref{casimir} we see that Casimir Scalling predicts a repulsive quark-quark and antiquark-antiquark interaction 
and an attractive quark-antiquark interaction for the state $|S\rangle$. As can be seen in Figs. \ref{tetraq_n1},
\ref{tetraq6_cuts} and \ref{tetraq8_cuts}, the results qualitatively agree with this prediction, due to a suppression
of the flux-tube between the two quarks and the two antiquarks.
For the state $| \bar{I} \rangle$, the prediction is the repulsion between the particles that form the two mesons in the
ground states, while all other interactions are attractive. Again, the results do not qualitatively contradict this
possibility, as shown in Fig. \ref{tetraq_n1}, for $(r_1,r_2) = (6,8)$ and $(r_1,r_2) = (8,10)$, where we again
see a flux tube suppression between the particles that form mesons in the ground state.

So, the results for the first excited state seem to agree with the hypothesis that the first excited state is orthogonal
to the ground state in the color space, with $|\bar{I}\rangle$ and $|\bar{II}\rangle$ and $|S\rangle$ being the
excited states, where $|I\rangle$, $|II\rangle$ and $|A\rangle$ respectively are the ground states.

\section{Conclusion}

In this work, we use a variational method to compute the chromo-fields of the system composed of
two quarks and two antiquarks.
With this method we can not only observe the region where the tetraquark state is the ground one,
but also the region where the ground state is composed of two mesons, as well as the transitions between this
two regimes and the onw between the two possible meson-meson ground states.
We are also able to observe, for the first time, the first excited state of this system.

For the ground state the results improve our previous work \cite{Cardoso:2011fq} where only the tetraquark operator
was used.
There we had difficulties with the measurement of the ground state outside of the tetraquark region due to the
low overlap of the used operator with the true ground state. Here, we are able to overcome that difficulty by using
a variational basis. The results are similar, in the tetraquark region, to those obtained there, while giving
the expected transition to the meson-meson behavior outside that region.

The results for the first excited state are not, at the first view, as understandable as the ones for the ground state.
We note however that since the ground state is well bellow the first gluonic excitation of the string, the only way by which
we can explain this excitation is the flux tube recombination. This way we know that the first excited state has a color
eigenfunction which is orthogonal to the ground state, be it a tetraquark or a two-meson state.
By using, the Casimir factors, we compare the predictions of the aforementioned hypothesis with our results of the
Lagrangian density for this state. We conclude that both results are compatible.

Note that this first excited state should be as important as the ground state of this system.
Think for instance in the decay of a tetraquark into a two meson system. Since the color structure of the initial and
final states are not the same, we can not consider the potential of the system to be a scalar in the color space.
Instead it should be given by a two by two matrix as explainded above. To reconstruct this matrix we need to know
not only the potential of the ground state but also it's color composition as well as the potential of the first
excited state.

This study should be complemented in the future by a detailed study of the potential and color composition of the
first excited state in the lattice.

\begin{table}
\begin{tabular}{|c|c|c|c|}
\hline 
$|\Psi\rangle$ & $C_{12}$ & $C_{13}$ & $C_{14}$\tabularnewline
\hline 
\hline 
$|I\rangle$ & 0 & 1 & 0\tabularnewline
\hline 
$|\bar{I}\rangle$ & $\frac{1}{4}$ & $-\frac{1}{8}$ & $\frac{7}{8}$\tabularnewline
\hline 
$|II\rangle$ & 0 & 0 & 1\tabularnewline
\hline 
$|\bar{II}\rangle$ & $\frac{1}{4}$ & $\frac{7}{8}$ & $-\frac{1}{8}$\tabularnewline
\hline 
$|A\rangle$ & $\frac{1}{2}$ & $\frac{1}{4}$ & $\frac{1}{4}$\tabularnewline
\hline 
$|S\rangle$ & $-\frac{1}{4}$ & $\frac{5}{8}$ & $\frac{5}{8}$\tabularnewline
\hline 
\end{tabular}\caption{$C_{ij}=\langle\Psi|\frac{\lambda_{i}\cdot\lambda_{j}}{-16/3}|\Psi\rangle$}
\label{casimir}
\end{table}

\section{Acknowledgements}

This work was partly funded by the FCT contracts, PTDC/FIS/100968/2008 and CERN/FP/116383/2010.
Marco Cardoso is supported by FCT under the contract SFRH/BPD/73140/2010.
Nuno Cardoso is also supported by FCT under the contract SFRH/BD/44416/2008.

\bibliographystyle{apsrev4-1}
\bibliography{bib}

\end{document}